\documentclass[trackchanges,twocolumn]{aastex7}
\usepackage{graphicx}
\usepackage{amsmath}
\usepackage{booktabs}
\usepackage{CJK}

\begin{document}
\begin{CJK*}{UTF8}{gbsn}

\title{Probing Red Giant Interiors with g-Dominated Mixed Modes \uppercase\expandafter{\romannumeral2}: The Cases of KIC 6117517, KIC 9475697, KIC 10123207 and KIC 10200377}

\correspondingauthor{Xinyi Zhang, Shaolan Bi}
\email{zhangxy\_cn@bnu.edu.cn,bisl@bnu.edu.cn}

\author[0000-0003-1860-1851]{Xinyi Zhang} 
\affiliation{Institute for Frontiers in Astronomy and Astrophysics,\\
 Beijing Normal University, \\
 Beijing 102206, People{'}s Republic of China}
\affiliation{School of Physics and Astronomy,
 Beijing Normal University, Beijing 100875,
 People's Republic of China}
 \email{zhangxy\_cn@bnu.edu.cn} 

\author[0000-0001-6396-2563] {Tanda Li}
\affiliation{Institute for Frontiers in Astronomy and Astrophysics,\\
 Beijing Normal University, \\
 Beijing 102206, People{'}s Republic of China} 
\affiliation{School of Physics and Astronomy,
 Beijing Normal University, Beijing 100875,
 People's Republic of China}
\email{litanda@bnu.edu.cn}

\author[0000-0002-7642-7583] {Shaolan Bi} 
\affiliation{Institute for Frontiers in Astronomy and Astrophysics,\\
 Beijing Normal University, \\
 Beijing 102206, People{'}s Republic of China} 
\affiliation{School of Physics and Astronomy,
 Beijing Normal University, Beijing 100875,
 People's Republic of China}
\email{bisl@bnu.edu.cn}

\begin{abstract}
Building on Paper~I, which constrained core overshooting to $f_{\rm ov} < 0.01$ for $M < 1.4\,M_{\odot}$ and found $\alpha_{\rm MLT} \simeq 2.0$ for metal-rich red giants, we extend this analysis to four lower-mass \textit{Kepler} red 
giants spanning a broad metallicity range: KIC~6117517, KIC~9475697, KIC~10123207, and KIC~10200377. Two complementary 
model grids are constructed: Grid~A fixes $\alpha_{\rm MLT} = 2.0$ while varying metallicity; Grid~B adopts spectroscopic 
metallicities with $\alpha_{\rm MLT}$ as a free parameter. Grid~A successfully reproduces both asteroseismic and spectroscopic constraints for the metal-rich targets (KIC~6117517 and KIC~9475697), confirming $\alpha_{\rm MLT} = 2.0$ 
for metal-rich red giants. For the metal-poor targets (KIC~10123207 and KIC~10200377), Grid~B yields larger mixing-length parameters of $\alpha_{\rm MLT} \approx 2.16$--$2.18$. Although only four stars are analyzed with unequal sensitivity 
to $\alpha_{\rm MLT}$, these results suggest a tentative increase in convective efficiency toward lower metallicity, consistent with 3D hydrodynamical simulations but opposite to some 1D empirical calibrations. The best-fitting overshooting parameters ($f_{\rm ov} = 0.006$--$0.007$) independently support the Paper~I constraint and are consistent with a mass-dependent overshooting relation. The parameter uncertainties reach $0.6$--$2.5\%$ in mass, $0.2$--$0.9\%$ in radius, $2$--$13\%$ in age, and below $0.6\%$ in helium-core mass and below $0.2\%$ in helium-core radius. Surface-effect corrections are negligible for g-dominated mixed modes, validating $\chi^2_{l=1g}$ as a robust fitting metric. For KIC~9475697, the core rotation rate is determined to be $\Omega_{\rm core} = 0.6945 \pm 0.0190\,\mu$Hz from model-based trapping parameters.
\end{abstract}

\keywords{\uat{Asteroseismology}{73} --- \uat{Stellar oscillations}{1617} --- \uat{Stellar ages}{1581} --- \uat{Red giant stars}{1372} }


\section{Introduction} \label{sec1}

Red giants are valuable tracers of stellar evolution and Galactic structure owing to their high luminosity and well-characterized oscillation properties. High-precision space photometry has revealed that red giants exhibit solar-like oscillations driven by 
near-surface convection \citep{2009Natur.459..398D}, including mixed modes formed by the coupling of gravity waves in the radiative core with acoustic waves in the convective envelope \citep{2009A&A...506...57D}. The period spacings of mixed modes distinguish hydrogen-shell-burning red-giant-branch (RGB) stars from core-helium-burning stars \citep{2011Natur.471..608B,2011A&A...532A..86M,2014A&A...572L...5M}, while individual oscillation frequencies are sensitive to the detailed internal density and sound-speed profiles throughout the stellar interior, enabling tight constraints on fundamental 
stellar parameters such as mass, radius, age, and helium-core structure \citep{2013ARA&A..51..353C,2021RvMP...93a5001A}. 
Rotational splittings of mixed modes additionally reveal strong radial differential rotation in evolved stars, with stellar cores rotating far more rapidly than their envelopes \citep{2012Natur.481...55B,2012ApJ...756...19D,2012A&A...548A..10M}. Detailed asteroseismic modeling of \textit{Kepler} red giants has therefore achieved increasingly precise determinations of their global 
and core properties, with typical precisions of $\sim$5--10\% in mass and radius and $\sim$10--30\% in age \citep{perez2016,2017ApJ...844..102H,2018ApJ...855...16Z,2020MNRAS.494..511Z,2021MNRAS.505.2336M,2022ApJ...931...64Z,2023ApJ...953..182W,2026ApJ..1002..198Z}.

Stellar models of this kind, however, still carry uncertainties arising from poorly constrained input physics, which can introduce systematic errors into the inferred stellar parameters. Two such ingredients are particularly relevant here: the mixing-length 
parameter $\alpha_{\rm MLT}$, which describes convection in one-dimensional models, and the convective-core overshooting parameter $f_{\rm ov}$, which governs the extent of extra mixing beyond the formal convective boundary during the main sequence. The appropriate value of $\alpha_{\rm MLT}$ and its dependence on stellar parameters remain subjects of active debate: several empirical calibrations find a positive correlation between $\alpha_{\rm MLT}$ and metallicity \citep{2014ApJS..214...27M,2017A&A...601A..67C,2018ApJ...858...28V,2024ApJ...974...77L}, whereas three-dimensional hydrodynamical simulations predict a weak or inverse dependence \citep{2015A&A...573A..89M}. Regarding the value of $\alpha_{\rm MLT}$ in red giants specifically, \citet{2018MNRAS.475..981L} reported that red giants generally require larger mixing-length values than 
the Sun. The overshooting parameter $f_{\rm ov}$ significantly influences the helium-core mass formed during the main sequence and, in turn, affects stellar structure and age estimation during the RGB phase \citep{2013sse..book.....K}. Our preceding 
study \citep[hereafter Paper~I]{2026ApJ..1002..198Z}, which modeled four \textit{Kepler} red giants by fully exploiting g-dominated mixed modes, found that $\alpha_{\rm MLT} \simeq 2.0$ provided satisfactory fits for all targets, and further revealed that lower-mass red giants ($M \lesssim 1.4\,M_{\odot}$) tend to favor $f_{\rm ov} < 0.01$, while more massive stars require larger overshooting values. 

Based on these results, the present work extends the sample to four additional \textit{Kepler} RGB stars spanning a wider metallicity baseline, with the aim of further examining the role of $\alpha_{\rm MLT}$ and consolidating the $f_{\rm ov}$--mass trend identified in Paper~I. Specifically, we compare a grid with fixed $\alpha_{\rm MLT} = 2.0$ against a grid in which 
$\alpha_{\rm MLT}$ is freely varied for every star, and we quantify the sensitivity of the inferred global and helium-core properties to this choice. The overshooting range $0.005 \leq f_{\rm ov} \leq 0.010$ is adopted from Paper~I as an empirical prior 
consistent with the low-mass regime of our sample ($M \lesssim 1.4\,M_{\odot}$); any $\alpha_{\rm MLT}$--metallicity behavior inferred from the present four stars is treated as an exploratory trend that motivates future work with broader samples. In addition, we exploit the rotational multiplets identified in KIC~9475697 to constrain its internal rotation profile, following the same approach as in Paper~I. The rest of this paper is organized as follows: photometric and spectroscopic data are presented in 
Section~\ref{sec2}. The input physics, model grids, and fitting method are described in Sections~\ref{sec3.1}, \ref{sec3.2}, and \ref{sec3.3}, respectively. The fitting results of Grid~A are presented in Section~\ref{sec4.1}. The mixing-length parameter $\alpha_{\rm MLT}$ is discussed in Section~\ref{sec4.2}. The selection of the best-fitting model is discussed in Section~\ref{sec4.3}. The overshooting parameter $f_{\rm ov}$ is discussed in Section~\ref{sec4.4}. The envelope and core rotation of KIC~9475697 are presented in Section~\ref{sec4.5}. Our conclusions are summarized in Section~\ref{sec5}.


\section{Photometric and Spectroscopic Observations} \label{sec2}

The \textit{Kepler} mission continuously observed KIC~6117517, KIC~9475697, KIC~10123207, and KIC~10200377 for four years, yielding high-precision photometric data. The individual oscillation frequencies for these targets are adopted from \citet{corsaro2015}. We selected these stars because their seismic properties place them in the low-mass RGB regime ($M\simeq1.1$--$1.3\,M_{\odot}$), while their spectroscopic metallicities span approximately $-0.6\lesssim {\rm [Fe/H]}\lesssim+0.2$. This combination complements Paper~I by expanding the sample below $1.4\,M_{\odot}$ (where Paper~I suggested $f_{\rm ov}<0.01$) and by providing a sufficiently broad metallicity range to test whether the $\alpha_{\rm MLT}\simeq2.0$ calibration is universal or metallicity-dependent. Furthermore, their rich, well-resolved $l=1$ mixed-mode spectra enable a homogeneous modeling analysis across the entire sample. The asteroseismic and spectroscopic parameters for these four red giants are summarized in Table~\ref{t1}. All four stars exhibit mixed modes, and their individual oscillation frequencies are provided in Tables~\ref{ta1}--\ref{ta4}. The parameters $\Delta\nu$ and $\Delta\Pi_{\rm 1,obs}$ listed in Table~\ref{t1} represent the observed large frequency separation and asymptotic period spacing, respectively.

\begin{deluxetable*}{lcccc}
\tablecaption{Stellar Parameters of Four Red Giants}
\label{t1}
\tablehead{
\colhead{Parameter} & \colhead{KIC 6117517} & \colhead{KIC 9475697} & \colhead{KIC 10123207} & \colhead{KIC 10200377}
}
\startdata
$T_{\rm eff}$ (K) &$4734\pm91^{a}$          &$4791\pm91^{a}$          &$4840\pm91^{a}$          &$4828\pm91^{a}$ \\
                  &$4673.806\pm7.001^{b}$   &$4769.4805\pm7.8821^{b}$ &...                      &$4882.346\pm14.573^{b}$ \\
                  &$4671.26\pm44.41^{c}$    &...                      &...                      &$4958.82\pm17.23^{c}$   \\
                  &$4630.36\pm35.80^{c}$    &...                      &...                      &$4988.41\pm23.52^{c}$\\
                  &$4620.43\pm28.52^{c}$    &...                      &...                      &$4952.09\pm154.87^{c}$\\
                  &...                      &...                      &...                      &$4985.04\pm37.33^{c}$\\
log $g$ (dex)     &$3.01\pm0.11^{a}$        &$2.90\pm0.11^{a}$        &$2.98\pm0.11^{a}$        &$3.00\pm0.11^{a}$ \\
                  &$2.923\pm0.019^{b}$      &$2.954\pm0.0210^{b}$     &...                      &$3.113\pm0.036^{b}$ \\
                  &$3.131\pm0.073^{c}$      &...                      &...                      &$3.130\pm0.028^{c}$ \\
                  &$3.206\pm0.050^{c}$      &...                      &...                      &$3.267\pm0.039^{c}$  \\
                  &$3.136\pm0.042^{c}$      &...                      &...                      &$3.231\pm0.243^{c}$  \\
                  &...                      &...                      &...                      &$3.220\pm0.062^{c}$   \\
$\rm [Fe/H]$      &$0.38\pm0.03^{a}$        &$0.19\pm0.03^{a}$        &$-0.45\pm0.04^{a}$       &$-0.63\pm0.04^{a}$ \\
                  &$0.2224\pm0.0067^{b}$    &$0.0456\pm0.0069^{b}$    &...                      &$-0.5812\pm0.0115^{b}$ \\
                  &$0.243\pm0.040^{c}$      &...                      &...                      &$-0.553\pm0.015^{c}$\\
                  &$0.213\pm0.029^{c}$      &...                      &...                      &$-0.547\pm0.021^{c}$\\
                  &$ 0.212\pm0.023^{c}$     &...                      &...                      &$-0.526\pm0.144^{c}$\\
                  &...                      &...                      &...                      &$-0.520\pm0.035^{c}$\\
$\Delta\nu$ ($\mu$Hz)       &$10.16\pm0.05^{a}$   &$9.88\pm0.05^{a}$    &$13.67\pm0.07^{a}$   &$12.47\pm0.06^{a}$ \\
$\nu_{\rm max}$ ($\mu$Hz)   &$120.3\pm0.2^{a}$    &$115.1\pm0.2^{a}$    &$160.9\pm0.2^{a}$    &$142.5\pm0.2^{a}$ \\
$\Delta\Pi_{\rm 1,obs}$ (s)        &$76.91\pm0.02^{d}$   &$75.70\pm0.04^{d}$   &$83.88\pm0.03^{d}$   &$81.58\pm0.03^{d}$ \\
\enddata
\tablenotetext{a}{\citet{perez2016}}
\tablenotetext{b}{\citet{2022ApJS..259...35A}}
\tablenotetext{c}{\citet{2018ApJS..238...30Z}}
\tablenotetext{d}{\citet{Mosser2012}}
\end{deluxetable*}

\section{Asteroseismic Modeling} \label{sec3}
\subsection{Input Parameters} \label{sec3.1}
We compute one-dimensional stellar-evolution models with Modules for Experiments in Stellar Astrophysics (MESA; \citealp{2011ApJS..192....3P,2013ApJS..208....4P,2015ApJS..220...15P,2018ApJS..234...34P}), version 10398. Adiabatic oscillation frequencies are calculated with ADIPLS \citep{2008Ap&SS.316..113C} through MESA's ``pulse\_adipls'' interface. The calculations do not account for semiconvection, rotation, or magnetic fields. The OPAL EOS tables published by \citet{2002ApJ...576.1064R} are used for the equation of state (EOS) calculations, while \citet{1996ApJ...464..943I} and \citet{2005ApJ...623..585F} are used for the high- and low-temperature regions of the opacity tables, respectively. The simple atmospheric model is used to calculate the surface temperature and pressure. The initial metal mixture follows \citet{1998SSRv...85..161G} (GS98). We adopt the mixing-length theory (MLT) of \citet{1958ZA.....46..108B} to deal with stellar convection. Convective-core overshooting is treated using the prescription of \citet{2000A&A...360..952H}. The overshoot mixing diffusion coefficient $D_{\rm ov}$ is defined as
\begin{equation}
D_{\rm ov}=D_{\rm conv,0}\exp\left (\frac{-2z}{f_{\rm ov}H_{\rm p,0}} \right ),
\label{e1}
\end{equation}
where $f_{\rm ov}$ is an adjustable parameter, $D_{\rm conv,0}$ is the MLT-derived diffusion coefficient at a user-defined location near the Schwarzschild boundary, $H_{\rm p,0}$ is the pressure scale height at
that location, and $z$ is the distance in the radiative layer away from that location (for more details, see \citealp{2000A&A...360..952H}; \citealp{2011ApJS..192....3P}).

\subsection{Model Grids}\label{sec3.2}

Table~\ref{t2} lists the input parameters for KIC~6117517, KIC~9475697, KIC~10123207, and KIC~10200377. Each model is evolved from the pre-main sequence to the red-giant stage. For all four stars, the initial helium mass fraction follows $Y=0.245+1.54Z$ \citep{2008ApJS..178...89D,2014AJ....148...85T}. Because the adopted mass ranges lie below $1.4\,M_{\odot}$, we use the Paper~I low-mass constraint as an empirical prior and set $f_{\rm ov}=0.005$--$0.010$ in both grids. Our previous modeling of KIC~9145955 and KIC~9970396 found that their g-dominated mixed modes favored $\alpha_{\rm MLT}\simeq2.0$ \citep{2018ApJ...855...16Z,2020MNRAS.494..511Z}. Based on that calibration, Grid~A fixes $\alpha_{\rm MLT}=2.0$ while varying $Z$. Grid~B instead fixes the metallicity of each star at its adopted spectroscopic value while varying $\alpha_{\rm MLT}$. The two grids are analyzed in Sections~\ref{sec4.1} and \ref{sec4.2}.

\begin{table*}
	\caption{\label{t2}
	Initial Input Parameters of Theoretical Models.}
	\centering
	\begin{tabular}{ccccc}
	\hline\hline
	&mass M/$\rm \delta M$ 	&metallicity Z/$\rm \delta Z$ 	 &Mixing-length parameter $\alpha_{\rm MLT}$\ /$\delta \alpha$ 
    &core overshooting parameter $f_{ov}$ / $\delta f_{ov}$\\
				&($\rm M_{\odot}$)\\
	\hline
    &&&Grid A\\
	KIC 6117517	    &1.19-1.30/0.01	    &0.022-0.026/0.001	   	&2.0		   	&0.005-0.010/0.001\\
	KIC 9475697		&1.15-1.35/0.01		&0.015-0.019/0.001		&2.0			&0.005-0.010/0.001\\
	KIC 10123207	&1.10-1.21/0.01		&0.003-0.008/0.001		&2.0			&0.005-0.010/0.001\\
	KIC 10200377  	&1.05-1.26/0.01		&0.001-0.006/0.001      &2.0       	    &0.005-0.010/0.001\\
	\hline
    &&&Grid B\\
    KIC 6117517	    &1.19-1.30/0.01	    &0.025	   	&1.8-2.3/0.1		   	&0.005-0.010/0.001\\
	KIC 9475697		&1.15-1.35/0.01		&0.017		&1.8-2.3/0.1			&0.005-0.010/0.001\\
	KIC 10123207	&1.10-1.21/0.01		&0.005		&1.8-2.3/0.1			&0.005-0.010/0.001\\
	KIC 10200377  	&1.05-1.26/0.01		&0.003      &1.8-2.3/0.1      	    &0.005-0.010/0.001\\
    \hline
	\end{tabular}
\end{table*}

\subsection{Model Fitting}\label{sec3.3}
In our previous work \citep{2018ApJ...855...16Z,2020MNRAS.494..511Z}, we developed a method that uses observed and theoretical frequencies to distinguish the most p-dominated mixed mode in each radial order from the accompanying g-dominated mixed modes. This classification enables consistent matching of the observed and modeled g-dominated modes.

In this work, we apply this identification to the $l=1$ frequencies of KIC~6117517, KIC~9475697, KIC~10123207, and KIC~10200377. In red giants, the observed g-dominated mixed modes occur predominantly at $l=1$. Although $l=2$ modes are also mixed, only those close to pure acoustic modes are generally detectable \citep{2017A&ARv..25....1H}. For each modeled radial order, we define the mixed mode with the lowest inertia as the most p-dominated mode and classify the remaining modes as g-dominated. The normalized mode inertia is \citep{2010aste.book.....A,2017A&ARv..25....1H}

\begin{equation}
I=\frac{4\pi\int_{0}^{R}[\xi_{r}(r)^{2}+l(l+1)\xi_{h}(r)^{2}]\rho_{0}r^{2}dr}{M[\xi_{r}(R_{\rm phot})^{2}+l(l+1)\xi_{h}(R_{\rm phot})^{2}]},
\label{e2}
\end{equation}

For the observations, we identify the $l=1$ mixed mode closest to the expected pure $l=1$ p-mode frequency in each radial order as the most p-dominated mode. Further details are given in \citet{2020MNRAS.494..511Z} and Paper~I.

We employ two goodness-of-fit statistics to evaluate model agreement with the observed oscillation frequencies. The first is a $\chi^2$-type function normalized by the observational uncertainties:
\begin{equation}
\chi^{2}_{\rm all}=\frac{1}{K_{\rm all}}\sum_{i=1}^{K_{\rm all}}
\left(\frac{\nu_{i(\rm all)}^{\rm obs}-\nu_{i(\rm all)}^{\rm mod}}
{\sigma^{\rm obs}_{i(\rm all)}}\right)^2,
\label{e3}
\end{equation}

\begin{equation}
\chi^{2}_{l=1g}=\frac{1}{K_{l=1g}}\sum_{i=1}^{K_{l=1g}}
\left(\frac{\nu_{i(l=1g)}^{\rm obs}-\nu_{i(l=1g)}^{\rm mod}}
{\sigma^{\rm obs}_{i(l=1g)}}\right)^2,
\label{e4}
\end{equation}

where superscripts ``obs'' and ``mod'' denote observed and model frequencies, respectively; subscripts ``all'' and ``$l=1g$'' indicate that all observed frequencies or only the $l=1$ g-dominated mixed-mode frequencies are included; $\sigma^{\rm obs}$ is the observational uncertainty of each frequency; and $K_{\rm all}$ and $K_{l=1g}$ are the corresponding total mode counts. 
For each evolutionary track, the optimal model is the one with the minimum $\chi^{2}_{l=1g}$, denoted $\chi^{2}_{m,l=1g}$.
The second statistic is an uncertainty-independent mean squared residual:

\begin{equation}
S^{2}_{\rm all}=\frac{1}{K_{\rm all}}\sum_{i=1}^{K_{\rm all}}
\left(\nu_{i(\rm all)}^{\rm obs}-\nu_{i(\rm all)}^{\rm mod}
\right)^2,
\label{e5}
\end{equation}

\begin{equation}
S^{2}_{l=1g}=\frac{1}{K_{l=1g}}\sum_{i=1}^{K_{l=1g}}
\left(\nu_{i(l=1g)}^{\rm obs}-\nu_{i(l=1g)}^{\rm mod}
\right)^2,
\label{e6}
\end{equation}

where the symbols follow the same definitions as in equations~(\ref{e3})--(\ref{e4}). Because $S^2$ is not normalized by the observational uncertainties, its square root $\sqrt{S^{2}}$ directly measures the absolute rms frequency residual in $\mu$Hz.

\section{Results} \label{sec4}

\subsection{Grid~A Fitting Results}\label{sec4.1}
\begin{figure*}[t]
\centering
\includegraphics[angle=0,width=\textwidth]{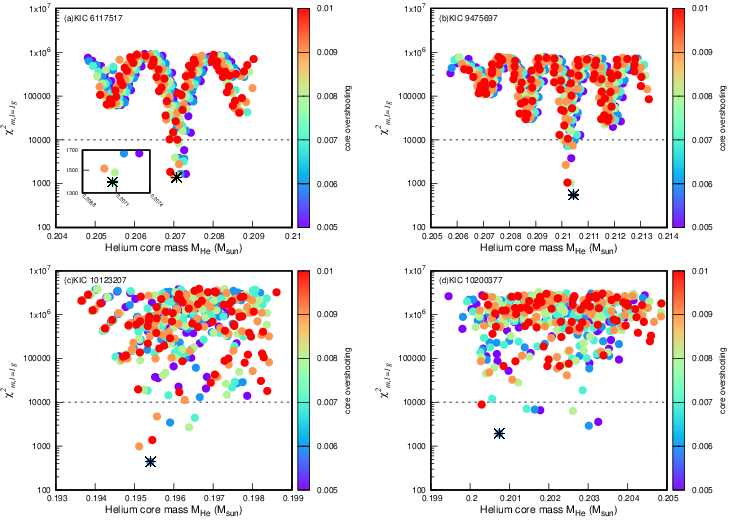}
\caption{$\chi^{2}_{m,l=1g}$ as a function of helium-core mass $M_{\rm He}$ for KIC~6117517 (panel a), KIC~9475697 (panel b), KIC~10123207 (panel c), and KIC~10200377 (panel d). Each circle represents the model with the minimum $\chi^{2}_{l=1g}$ on an evolutionary track. Black asterisks mark the absolute minima among the plotted Grid~A tracks. The inset in panel (a) enlarges the region around the absolute minimum, where several models have closely spaced values of $\chi^{2}_{m,l=1g}$.}
\label{f1}
\end{figure*}

Figure~\ref{f1} shows $\chi^{2}_{m,l=1g}$ as a function of helium core mass $M_{\rm He}$ for all four targets, where each circle represents the model with the minimum $\chi^{2}_{l=1g}$ on each evolutionary track, and the color encodes the core overshooting parameter $f_{\rm ov}$. The absolute minima of $\chi^{2}_{m,l=1g}$ among the plotted Grid~A evolutionary tracks are approximately 1392, 556, 438, and 1944 in panels (a)--(d), respectively; these points are marked by black asterisks in Figure~\ref{f1}. The inset in panel (a) enlarges the region around the absolute minimum, where several models have similar values of $\chi^{2}_{m,l=1g}$. The dashed horizontal line at $\chi^{2}_{m,l=1g} = 10{,}000$ marks the threshold used to define the candidate model sample for subsequent best-fitting model selection (Section~\ref{sec4.3}). The large absolute $\chi^{2}$ values reflect the exceptional frequency precision of \textit{Kepler} photometry rather than a failure of the stellar models. As listed in Table~\ref{t3}, the mean observational uncertainty is $\bar{\sigma}_{\rm obs} \sim 0.003\,\mu$Hz, while the rms fitting residuals of the best-fitting models are $\sqrt{S^{2}_{m,l=1g}} \sim 0.03$--$0.05\,\mu$Hz, roughly one order of magnitude larger. Because each term in $\chi^{2}_{m,l=1g}$ is normalized by $(\sigma^{\rm obs}_i)^2$, this mismatch alone inflates $\chi^{2}_{m,l=1g}$ to $\sim 10^{2}$--$10^{3}$, fully consistent with the values seen in Figure~\ref{f1}. Comparable behavior was reported in Paper~I for a similar sample of \textit{Kepler} red giants, where the same grid resolution yielded $\sqrt{S^{2}_{m,l=1g}} \sim 0.04$--$0.06\,\mu$Hz and yet the helium core mass and radius were determined with high precision. The present residuals are fully consistent with those of Paper~I, confirming that the current grid resolution is sufficient 
for precise determination of the helium core properties. Following the procedure described in \citet{2026ApJ..1002..198Z}, we therefore adopt a rescaled likelihood function to obtain robust grid-based parameter estimates.
\begin{equation}
\mathcal{L}_i = \exp\left( -\frac{1}{2} 
\frac{\chi^2_i - \chi^2_{\mathrm{min}}}
{\chi^2_{\mathrm{min}}} \right),
\label{e7}
\end{equation}

The normalization by $\chi^{2}_{\min}$ in the denominator accounts for the case where $\chi^{2}_{\min} \gg 1$, which would otherwise cause all likelihoods to vanish numerically.
\begin{equation}
\langle p \rangle = \frac{\sum_{i=1}^{N} \mathcal{L}_i p_i}
{\sum_{i=1}^{N} \mathcal{L}_i},
\label{e8}
\end{equation}

\begin{equation}
 \sigma_p  = \left( \frac{\sum_{i=1}^{N} \mathcal{L}_i 
 (p_i - \langle p \rangle)^2}
 {\sum_{i=1}^{N} \mathcal{L}_i} \right)^{1/2},
\label{e9}
\end{equation}

where $\mathcal{L}$ is the likelihood function, $\chi^{2}_{\min}$ is the minimum $\chi^{2}$ value among all models, $p$ represents any fundamental stellar parameter of interest (e.g., mass, radius, age), $\langle p\rangle$ is the likelihood-weighted mean value of the fundamental parameters, $\sigma_{p}$ is the standard deviation, and $N$ is the total number of models in the grid. We use $\chi^{2}_{l=1g}$ to compute the likelihood weights. The resulting likelihood-weighted means and standard deviations for all four stars are listed in Table~\ref{t4}, where the mixing-length parameter is fixed at $\alpha_{\rm MLT} = 2.0$, a value calibrated for red giants in our previous work \citep{2018ApJ...855...16Z,2020MNRAS.494..511Z}. 

$\Delta\nu_{m}$ and $\Delta\Pi_{\rm 1, mod}$ in Table~\ref{t4} represent the model large frequency separation and the theoretical asymptotic period spacing, respectively. We adopt the method from \citet{2022ApJ...927..167L,2023ApJ...953..182W} to obtain the model large separation $\Delta\nu_{m}$:
\begin{equation}
\Delta\nu_{m} = \Delta\nu_{\odot}\left 
(\frac{\bar{\rho}}{\bar{\rho}_{\odot }} \right )^{0.507},
\label{e10}
\end{equation}
Here, we adopt the solar reference value $\Delta\nu_{\odot}=135.1\,\mu{\rm Hz}$ from \citet{2011ApJ...743..143H}. The quantities $\bar{\rho}_{\odot}$ and $\bar{\rho}$ are the mean solar density and mean stellar density, respectively. The theoretical asymptotic period spacing $\Delta\Pi_{1,\rm mod}$ is defined as
\citep{1980ApJS...43..469T,2010aste.book.....A}

\begin{equation}
\Delta\Pi_{1,mod}=\frac{2\pi^{2}}{\sqrt{2}\left 
(\int_{r1}^{r2}\frac{N}{r}dr \right )},
\label{e11}
\end{equation}
where $r_{1}$ and $r_{2}$ are the inner and outer boundaries of the region where gravity waves propagate, respectively.

We identify a single representative model for each star, referred to as the ``best-fitting model,'' for the subsequent \'{e}chelle-diagram comparisons. We select this model from the candidate pool (defined by $\chi^2_{m,l=1g}<10{,}000$) by requiring its surface gravity $\log g$ and metallicity $\rm [Fe/H]$ to be simultaneously consistent with the spectroscopic observations within their reported uncertainties. The model metallicity is calculated from the surface hydrogen ($X$) and metal ($Z$) abundances as \citep{2005essp.book.....S}
\begin{equation}
\rm [Fe/H]=\log\left(\frac{Z}{X}\right)_{\rm star}
-\log\left(\frac{Z}{X}\right)_{\odot},
\label{e12}
\end{equation}
where we adopt $(Z/X)_{\odot} = 0.0229$ \citep{1998SSRv...85..161G}. For the spectroscopic reference values, we adopt the APOGEE metallicities \citep{2022ApJS..259...35A} for stars with available APOGEE observations (KIC~6117517, KIC~9475697, and KIC~10200377), 
and the value from \citet{perez2016} for KIC~10123207, which lacks APOGEE observations.

The input parameters and derived properties of the best-fitting models from Grid~A (Models A--D) are listed in Table~\ref{t5}. For KIC~6117517 and KIC~9475697, the Grid~A best-fitting model metallicities ($\rm [Fe/H] = 0.224$ and $0.052$, respectively) agree with the APOGEE spectroscopic values to within $0.01$~dex, confirming that a fixed mixing-length parameter of $\alpha_{\rm MLT} = 2.0$ is appropriate for these two targets. For KIC~10123207 and KIC~10200377, however, the Grid~A metallicities ($\rm [Fe/H] = -0.381$ and $-0.542$) exceed the spectroscopic values by $0.07$~dex and $0.04$~dex, respectively, beyond the reported observational uncertainties. These discrepancies suggest that a fixed $\alpha_{\rm MLT} = 2.0$ is not appropriate for these more metal-poor targets, motivating the Grid~B analysis presented in Section~\ref{sec4.2}.

\begin{table*}
\centering
\caption{\label{t3} Mean observational frequency 
uncertainties ($\bar{\sigma}_{\rm obs}$) and rms fitting 
residuals ($\sqrt{S^2_{l=1g}}$) for the four targets.}
\begin{tabular}{ccc}
\hline\hline
Target 
  & $\bar{\sigma}_{\rm obs}$ ($\mu$Hz)
  & $\sqrt{S^{2}_{l=1g}}$ ($\mu$Hz) \\
\hline
KIC~6117517  & 0.0028 & 0.037 \\
KIC~9475697  & 0.0043 & 0.045 \\
KIC~10123207 & 0.0029 & 0.040 \\
KIC~10200377 & 0.0038 & 0.031 \\
\hline
\end{tabular}
\end{table*}

\begin{table*}
\caption{\label{t4}Likelihood-weighted mean values and 
standard deviations of the fundamental parameters for the 
four targets derived from Grid~A, with the mixing-length 
parameter fixed at $\alpha_{\rm MLT} = 2.0$.}
\centering
\begin{tabular}{lccccc}
\hline
\hline
Parameters  &KIC6117517 &KIC9475697 &KIC10123207 &KIC10200377 \\
\hline
Mass(M/$M_{\odot}$)          		
  & $1.255 \pm 0.008$  & $1.272 \pm 0.008$  
  & $1.142 \pm 0.011$  & $1.210 \pm 0.030$ \\
Metal fraction(Z)            		
  & $0.024 \pm 0.001$  & $0.0171 \pm 0.0002$  
  & $0.008 \pm 0.001$  & $0.005 \pm 0.001$ \\
Overshooting parameter($f_{ov}$)  	
  & $0.007 \pm 0.002$  & $0.008 \pm 0.002$  
  & $0.008 \pm 0.002$  & $0.006 \pm 0.001$ \\
Effective temperature($\rm T_{eff}$/K) 		
  & $4738.7 \pm 10.5$  & $4813.4 \pm 4.8$  
  & $5006.5 \pm 12.4$  & $5071.0 \pm 25.6$ \\
Surface gravity $\log g$ (dex)    				
  & $2.988 \pm 0.001$  & $2.974 \pm 0.001$  
  & $3.146 \pm 0.001$  & $3.102 \pm 0.003$ \\
Radius(R/$R_{\odot}$)        				
  & $5.945 \pm 0.013$  & $6.087 \pm 0.013$  
  & $4.731 \pm 0.015$  & $5.120 \pm 0.044$ \\
Luminosity($\rm L_{\odot}$)	    			
  & $16.01 \pm 0.13$   & $17.87 \pm 0.12$   
  & $12.63 \pm 0.17$   & $15.58 \pm 0.58$ \\
$\Delta\nu_{m}$ ($\mu$Hz) 
  & $10.072 \pm 0.007$  & $9.784 \pm 0.002$  
  & $13.592 \pm 0.006$  & $12.413 \pm 0.009$ \\
$\Delta\Pi_{1,mod}$ (s)    		
  & $76.921 \pm 0.012$  & $75.704 \pm 0.003$  
  & $83.793 \pm 0.010$  & $81.511 \pm 0.012$ \\
Age(t/Gyr)                           		
  & $5.688 \pm 0.128$  & $4.849 \pm 0.127$  
  & $5.451 \pm 0.222$  & $3.956 \pm 0.493$ \\
$M_{He}$ ($M_{\odot}$)   
  & $0.2071 \pm 0.0001$  & $0.2103 \pm 0.0001$  
  & $0.1954 \pm 0.0003$  & $0.2020 \pm 0.0011$ \\
$R_{He}$ ($R_{\odot}$) 
  & $0.03110 \pm 0.00006$  & $0.03070 \pm 0.00003$  
  & $0.03115 \pm 0.00005$  & $0.03098 \pm 0.00005$ \\
\hline
\end{tabular}
\end{table*}

\begin{table*}
\caption{\label{t5}Input and Fundamental Parameters of the best-fitting models in Grid~A. Models A--D correspond to KIC~6117517, KIC~9475697, KIC~10123207, and KIC~10200377, respectively.}
\centering
\begin{tabular}{lcccc}
\hline
\hline
Parameters                            &KIC6117517 &KIC9475697 &KIC10123207 &KIC10200377 \\
                                      &Model A    &Model B    &Model C     &Model D  \\
\hline
&&Input Parameters\\
\hline
Mass (M/$M_{\odot}$)                  &1.26       &1.28       &1.16         &1.24\\
Initial Metal fraction (Z)            &0.025      &0.017      &0.006        &0.004\\
Overshooting parameter ($f_{ov}$)     &0.007      &0.006      &0.008        &0.006 \\
Mixing length parameter ($\alpha_{\rm MLT}$)    &2.0        &2.0        &2.0          &2.0\\
\hline
&&$\chi^{2}$ Value\\
\hline
$\chi_{l=1g}^{2}$                                &1392.159   &555.807   &2688.212  &2924.781\\
$\chi_{all}^{2}$                                 &1092.425   &558.047   &1859.825  &2802.690\\
$S^{2}_{l=1g}$                                   &0.00138    &0.00198   &0.00156   &0.000957\\
$S^{2}_{all}$                                    &0.0382     &0.0167    &0.0125    &0.0633\\
\hline
&&Fundamental Parameters\\
\hline
Effective temperature ($\rm T_{eff}$/K)              &4733.6    &4817.1    &5050.0   &5096.4 \\
$\rm [Fe/H]_{mod}$                                   &0.224     &0.052     &-0.381   &-0.542\\
Surface gravity $\log g$ (dex)                       &2.989     &2.975     &3.147    &3.106\\
Radius (R/$R_{\odot}$)                               &5.951     &6.099     &4.762    &5.162 \\
Luminosity ($\rm L_{\odot}$)                         &15.98     &18.00     &13.25    &16.15\\
Large frequency separation ($\Delta\nu_{m}$/$\mu$Hz) &10.078    &9.786     &13.565   &12.410\\
Period spacing ($\Delta\Pi_{1,mod}$/s)               &76.932    &75.702    &83.755   &81.513\\
Age (t/Gyr)                                          &5.650     &4.717     &4.773    &3.474 \\
Mass of helium core ($M_{He}$/$M_{\odot}$)           &0.2071    &0.2104    &0.1964   &0.2030 \\
Radius of helium core ($R_{He}$/$R_{\odot}$)         &0.03113   &0.03073   &0.03114  &0.03102\\
\hline
\end{tabular}
\end{table*}

\subsection{The Mixing-Length Parameter}\label{sec4.2}

We use Grid~B (Table~\ref{t2}) to investigate how varying the mixing-length parameter $\alpha_{\rm MLT}$ affects the inferred stellar parameters. APOGEE's high spectral resolution ($R\sim22{,}500$; \citealp{2016AJ....151..144G}) provides precise metallicity measurements that anchor the Grid~B calculations. For KIC~10123207, which lacks APOGEE observations, we adopt the value from \citet{perez2016}. The metallicity of each Grid~B target is therefore fixed at its adopted spectroscopic value.

\begin{figure*}[t]
\centering
\includegraphics[angle=0,scale=0.9]{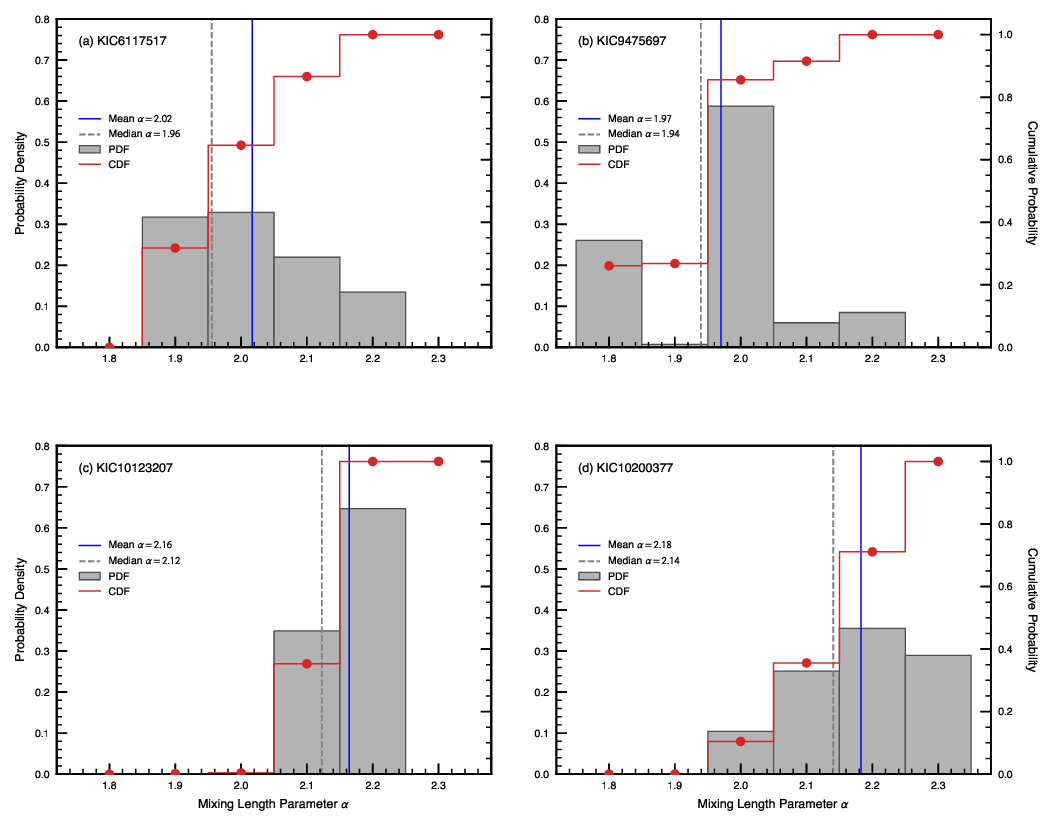}
\caption{Marginalized probability density functions (PDFs, grey bars) and cumulative distribution functions (CDFs, red stepped lines with filled circles) of the mixing-length parameter $\alpha_{\rm MLT}$, derived from $\chi^2_{l=1g}$-based likelihood weights, for KIC~6117517 (panel a), KIC~9475697 (panel b), KIC~10123207 (panel c), and KIC~10200377 (panel d). The blue vertical lines indicate the probability-weighted mean values, and the dashed grey lines represent the median values. }
\label{f2}
\end{figure*}

\begin{figure*}[t]
\centering
\includegraphics[angle=0,scale=1.2]{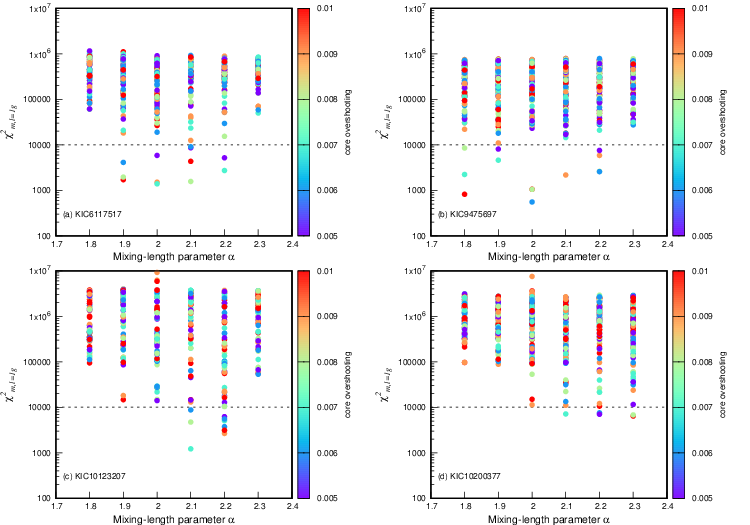}
\caption{$\chi^{2}_{m,l=1g}$ as a function of the mixing-length parameter $\alpha_{\rm MLT}$ for KIC~6117517 (panel a), KIC~9475697 (panel b), KIC~10123207 (panel c), and KIC~10200377 (panel d). Each circle represents the model with the minimum $\chi^{2}_{l=1g}$ on an evolutionary track, color-coded by the core overshooting parameter $f_{\rm ov}$. The horizontal dashed line marks the threshold $\chi^{2}_{m,l=1g}=10{,}000$ adopted to define the candidate model pool.}
\label{f3}
\end{figure*}

Figure~\ref{f2} shows the marginalized PDFs and CDFs of $\alpha_{\rm MLT}$ derived from $\chi^2_{l=1g}$-based likelihood weights. The four targets fall into two qualitatively distinct groups. KIC~10123207 exhibits a sharply peaked PDF concentrated at $\alpha_{\rm MLT} > 2.1$, yielding a well-constrained weighted mean of $2.16 \pm 0.05$. KIC~9475697 shows enhanced probability 
near $\alpha_{\rm MLT} \approx 1.9$--$2.0$ with a weighted mean of $1.97 \pm 0.12$, though the distribution retains a broad upper tail. By contrast, the PDFs of KIC~6117517 and KIC~10200377 are nearly flat across the sampled interval (Figure~\ref{f2}, panels a and d); their weighted means ($2.02 \pm 0.10$ and $2.18 \pm 0.10$, respectively) are sensitive to the adopted grid bounds and should not be interpreted as unique physical measurements. The CDFs, which give the cumulative posterior probability
$P(\alpha_{\rm MLT}\leq\alpha)$, provide a complementary view of the strength of these constraints. They reach a cumulative probability of 0.5 at median values of $\alpha_{\rm MLT}=1.96$, 1.94, 2.12, and 2.14 for KIC~6117517, KIC~9475697, KIC~10123207, and KIC~10200377, respectively. The relatively steep rises of the CDFs for KIC~9475697 and KIC~10123207 show that most of the
posterior probability is concentrated within comparatively narrow ranges of $\alpha_{\rm MLT}$. In contrast, the more gradual rises for KIC~6117517 and KIC~10200377 reflect their broader and less informative posterior distributions. The median values are
slightly lower than the corresponding weighted means, consistent with the modest tails toward higher $\alpha_{\rm MLT}$. The physical origin of this behavior is visible in Figure~\ref{f3}, which shows $\chi^2_{l=1g}$ as a function of $\alpha_{\rm MLT}$, color-coded by $f_{\rm ov}$. The two seismic observables that drive the fit, namely $\Delta\Pi_1$, which constrains the core structure through the combination of $M$ and $f_{\rm ov}$, and $\Delta\nu$, which constrains the mean density through $M$ and $\alpha_{\rm MLT}$, can in principle be reproduced by multiple parameter combinations. For KIC~6117517 and KIC~10200377, models  below the $\chi^2$ threshold (dashed line) are distributed across a wide range of $\alpha_{\rm MLT}$, with no single value strongly preferred; the color mixing confirms that different $f_{\rm ov}$ values participate in this degeneracy. For KIC~9475697 and KIC~10123207, no combination of $M$ and $f_{\rm ov}$ can fully compensate for frequency shifts induced by large excursions in $\alpha_{\rm MLT}$, so the minimum $\chi^2_{l=1g}$ is confined to a narrower interval, enabling a more informative constraint on convective efficiency.

\begin{figure*}[t]
\centering
\includegraphics[angle=0,scale=0.9]{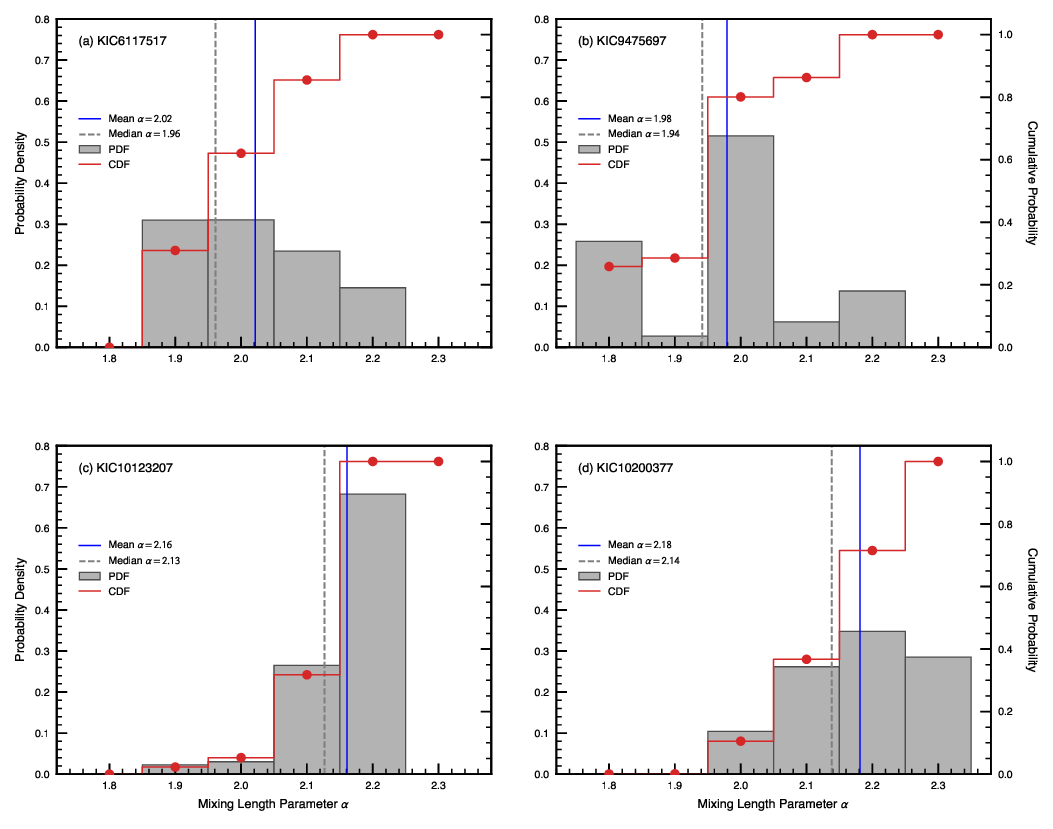}
\caption{Marginalized probability density functions (PDFs, grey bars) and cumulative distribution functions (CDFs, red stepped lines with filled circles) of the mixing-length parameter $\alpha_{\rm MLT}$, derived from $\chi^2_{all}$-based likelihood weights, for KIC~6117517 (panel a), KIC~9475697 (panel b), KIC~10123207 (panel c), and KIC~10200377 (panel d). The blue vertical lines indicate the probability-weighted mean values, and the dashed grey lines represent the median values.}
\label{f4}
\end{figure*}

\begin{figure*}[t]
\centering
\includegraphics[angle=0,scale=1.2]{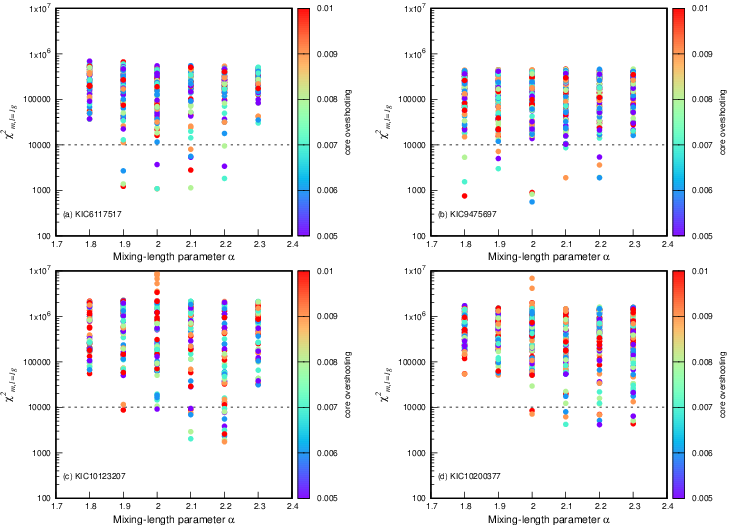}
\caption{$\chi^{2}_{m,\rm all}$ as a function of the mixing-length parameter $\alpha_{\rm MLT}$ for KIC~6117517 (panel a), KIC~9475697 (panel b), KIC~10123207 (panel c), and KIC~10200377 (panel d). Each circle represents the model with the minimum $\chi^{2}_{\rm all}$ on an evolutionary track, color-coded by the core overshooting parameter $f_{\rm ov}$. The horizontal dashed line marks the threshold $\chi^{2}_{m,\rm all}=10{,}000$ adopted to define the candidate model pool.}
\label{f5}
\end{figure*}

To assess whether the results depend on the choice of seismic constraints, we repeat the likelihood analysis using all matched oscillation frequencies ($\chi^2_{\rm all}$). The resulting PDFs, CDFs, and $\chi^2_{\rm all}$ diagrams are shown in Figures~\ref{f4} and \ref{f5}. As summarized in Table~\ref{t6}, the likelihood-weighted means and standard deviations of all fundamental parameters derived from $\chi^2_{l=1g}$ and $\chi^2_{\rm all}$ are consistent within the uncertainties, confirming that the results are robust regardless of whether g-dominated mixed modes alone or all available frequencies are used as constraints. This consistency reflects the dominant role of mixed modes in constraining red-giant stellar models. The two targets with informative $\alpha_{\rm MLT}$ posteriors point toward a possible metallicity dependence of convective efficiency. KIC~10123207  ($Z = 0.005$) requires $\alpha_{\rm MLT} \approx 2.16$, while KIC~9475697 ($Z = 0.017$) prefers $\alpha_{\rm MLT} \approx 1.97$. The remaining two stars are consistent with this direction but do not independently constrain it, given their broad posteriors. This tentative trend, where higher metallicity corresponds to lower convective efficiency, is qualitatively consistent with 3D hydrodynamical stellar atmosphere simulations \citep{2014MNRAS.445.4366T,2015A&A...573A..89M} and contrasts with at least one recent 1D empirical calibration based on eclipsing binaries \citep{2024ApJ...974...77L}.

Despite the variations in $\alpha_{\rm MLT}$ discussed above, the stellar parameters derived from Grid~A (Table~\ref{t4}) and Grid~B (Table~\ref{t6}) are generally consistent within the formal uncertainties when constrained by $\chi^2_{l=1g}$. In particular, the helium-core properties ($M_{\rm He}$ and $R_{\rm He}$), seismic parameters ($\Delta\nu_m$ and $\Delta\Pi_1$), and overshooting parameter ($f_{\rm ov}$) agree well between the two grids, indicating that the $l=1$ g-dominated mixed modes constrain these quantities with little sensitivity to the adopted $\alpha_{\rm MLT}$. We note that the age uncertainties are significantly larger in Grid~B due to the additional degree of freedom introduced by the free $\alpha_{\rm MLT}$ parameter. We therefore adopt the Grid~A results (Table~\ref{t4}) as our primary stellar parameters for the subsequent analysis.

\begin{table*}
\caption{\label{t6}Likelihood-weighted mean values and 
standard deviations of the fundamental parameters for the 
four targets derived from Grid~B, in which the metallicity 
$Z$ is fixed to the spectroscopic value and 
$\alpha_{\rm MLT}$ is a free parameter. For each star, the 
left sub-column gives results constrained by 
$\chi^{2}_{l=1g}$ and the right sub-column gives results 
constrained by $\chi^{2}_{\rm all}$.}
\centering
\resizebox{\textwidth}{!}{
\begin{tabular}{lcccccccc}
\hline\hline
 & \multicolumn{2}{c}{KIC~6117517} 
 & \multicolumn{2}{c}{KIC~9475697} 
 & \multicolumn{2}{c}{KIC~10123207} 
 & \multicolumn{2}{c}{KIC~10200377} \\
\cmidrule(lr){2-3}\cmidrule(lr){4-5}
\cmidrule(lr){6-7}\cmidrule(lr){8-9}
Parameters 
 & $\chi^{2}_{l=1g}$ & $\chi^{2}_{\rm all}$ 
 & $\chi^{2}_{l=1g}$ & $\chi^{2}_{\rm all}$ 
 & $\chi^{2}_{l=1g}$ & $\chi^{2}_{\rm all}$ 
 & $\chi^{2}_{l=1g}$ & $\chi^{2}_{\rm all}$ \\
\hline
Mass (M/$M_{\odot}$)          		
 & $1.249\pm0.034$ & $1.248\pm0.035$ 
 & $1.280\pm0.039$ & $1.277\pm0.043$ 
 & $1.134\pm0.019$ & $1.134\pm0.023$ 
 & $1.236\pm0.039$ & $1.236\pm0.039$ \\
Metal fraction ($Z$)            		
 & $0.025$ & $0.025$ 
 & $0.017$ & $0.017$ 
 & $0.005$ & $0.005$ 
 & $0.003$ & $0.003$ \\
$f_{ov}$  	
 & $0.008\pm0.001$ & $0.008\pm0.002$ 
 & $0.008\pm0.002$ & $0.008\pm0.002$ 
 & $0.008\pm0.001$ & $0.008\pm0.002$ 
 & $0.008\pm0.002$ & $0.008\pm0.002$ \\
$\alpha_{\rm MLT}$  	        
 & $2.02\pm0.10$ & $2.02\pm0.10$ 
 & $1.97\pm0.12$ & $1.98\pm0.13$ 
 & $2.16\pm0.05$ & $2.16\pm0.07$ 
 & $2.18\pm0.10$ & $2.18\pm0.10$ \\
$T_{\rm eff}$ (K) 		
 & $4738.5\pm48.0$ & $4740.7\pm48.7$ 
 & $4797.7\pm56.3$ & $4801.7\pm60.7$ 
 & $5131.0\pm14.3$ & $5129.4\pm21.0$ 
 & $5185.5\pm27.0$ & $5185.1\pm27.3$ \\
$\log g$ (cgs)    				
 & $2.988\pm0.003$ & $2.988\pm0.003$ 
 & $2.974\pm0.004$ & $2.974\pm0.005$ 
 & $3.144\pm0.003$ & $3.144\pm0.003$ 
 & $3.104\pm0.004$ & $3.104\pm0.004$ \\
Radius (R/$R_{\odot}$)        				
 & $5.935\pm0.058$ & $5.933\pm0.059$ 
 & $6.100\pm0.064$ & $6.096\pm0.072$ 
 & $4.724\pm0.025$ & $4.726\pm0.033$ 
 & $5.160\pm0.058$ & $5.161\pm0.057$ \\
Luminosity ($L_{\odot}$)	    			
 & $15.96\pm0.35$ & $15.98\pm0.36$ 
 & $17.71\pm0.48$ & $17.75\pm0.50$ 
 & $13.90\pm0.04$ & $13.89\pm0.09$ 
 & $17.29\pm0.28$ & $17.30\pm0.29$ \\
$\Delta\nu_{m}$ ($\mu$Hz) 
 & $10.075\pm0.010$ & $10.075\pm0.011$ 
 & $9.782\pm0.007$  & $9.782\pm0.008$ 
 & $13.573\pm0.012$ & $13.567\pm0.018$ 
 & $12.396\pm0.017$ & $12.395\pm0.017$ \\
$\Delta\Pi_{1}$ (s)    		
 & $76.928\pm0.013$ & $76.928\pm0.014$ 
 & $75.701\pm0.008$ & $75.701\pm0.010$ 
 & $83.762\pm0.013$ & $83.757\pm0.020$ 
 & $81.487\pm0.074$ & $81.485\pm0.100$ \\
Age (t/Gyr)                           		
 & $5.872\pm0.590$ & $5.891\pm0.602$ 
 & $4.771\pm0.516$ & $4.816\pm0.586$ 
 & $4.965\pm0.285$ & $4.972\pm0.333$ 
 & $3.339\pm0.374$ & $3.331\pm0.368$ \\
$M_{He}$ ($M_{\odot}$)   
 & $0.2070\pm0.0002$ & $0.2071\pm0.0002$ 
 & $0.2102\pm0.0003$ & $0.2103\pm0.0003$ 
 & $0.1971\pm0.0001$ & $0.1971\pm0.0002$ 
 & $0.2049\pm0.0007$ & $0.2049\pm0.0007$ \\
$R_{He}$ ($R_{\odot}$) 
 & $0.03113\pm0.00009$ & $0.03114\pm0.00009$ 
 & $0.03065\pm0.00011$ & $0.03066\pm0.00011$ 
 & $0.03120\pm0.00001$ & $0.03120\pm0.00003$ 
 & $0.03114\pm0.00006$ & $0.03113\pm0.00006$ \\
\hline
\end{tabular}}
\end{table*}

\subsection{Determination of the Best-fitting Models} \label{sec4.3}

Based on the Grid~A and Grid~B analyses presented in Sections~\ref{sec4.1} and \ref{sec4.2}, we adopt the Grid~A representative models for KIC~6117517 and KIC~9475697 (Models~A and B) and the Grid~B representatives for KIC~10123207 and KIC~10200377 (Models~E and F). The four models used in the subsequent figures are listed in Table~\ref{t7}.

The model surface gravities are in good agreement with independent constraints. For KIC~6117517 and KIC~9475697, our values ($\log g = 2.989$ and $2.975$) are consistent with the asteroseismic results of \citet{2023ApJ...953..182W} ($2.993^{+0.004}_{-0.005}$ and $2.973^{+0.004}_{-0.004}$, respectively) and with the available APOGEE spectroscopic measurements. For KIC~10200377, our value ($\log g = 3.107$) agrees with both APOGEE ($3.113 \pm 0.036$) and LAMOST 
($3.130 \pm 0.028$). For KIC~10123207, our value ($\log g = 3.142$) exceeds the spectroscopic estimate 
of \citet{perez2016} ($2.98 \pm 0.11$) by $\sim$0.16~dex; given that asteroseismic $\log g$ determinations are known to be more reliable than spectroscopic ones for red giants \citep{2011ApJ...730...63G}, we consider our value more trustworthy.

The model metallicities are generally consistent with the adopted spectroscopic references. For KIC~6117517 and KIC~9475697, the model values ($\rm [Fe/H] = +0.224$ and $+0.052$) agree well with the APOGEE measurements ($+0.222 \pm 0.007$ and $+0.046 \pm 0.007$), which we prioritize over other surveys owing to APOGEE's higher spectral resolution 
($R \sim 22{,}500$; \citealt{2016AJ....151..144G}). For KIC~10123207, the Model~E metallicity ($\rm [Fe/H] = -0.441$) is consistent with \citet{perez2016} ($-0.45 \pm 0.04$). For KIC~10200377, Models~D and F bracket the available spectroscopic values: Model~D gives $\rm [Fe/H] = -0.542$, consistent with the LAMOST value ($-0.553 \pm 0.015$), while Model~F gives 
$-0.647$, consistent with \citet{perez2016} ($-0.63 \pm 0.04$) but below the APOGEE measurement ($-0.581 \pm 0.012$). This suggests that the true metallicity of KIC~10200377 lies in the range $Z = 0.003$--$0.004$.

\begin{figure*}[t]
\centering
\includegraphics[angle=0,width=\textwidth]{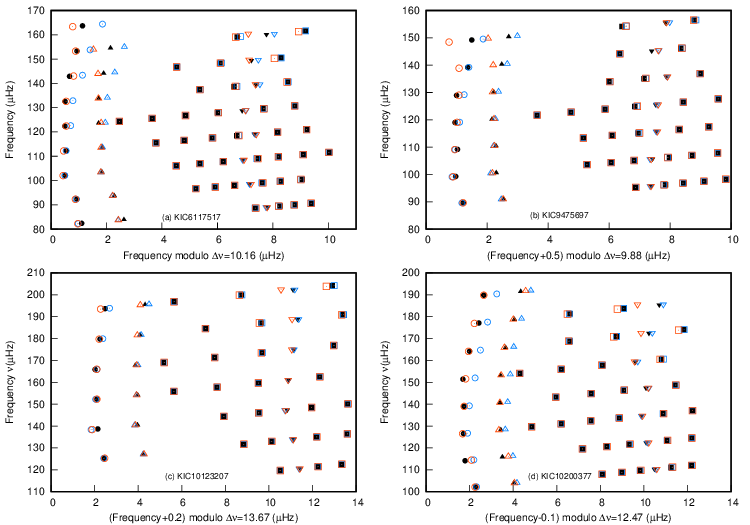}
\caption{Frequency \'{e}chelle diagrams of the best-fitting models for KIC~6117517, KIC~9475697, KIC~10123207, and KIC~10200377 (Models~A, B, E, and F in Table~\ref{t7}, respectively). Black filled symbols denote the observed frequencies, whereas blue and red open symbols denote the uncorrected and surface-corrected model frequencies, respectively. Triangles and circles denote $l=0$ and $l=2$ modes, respectively. Squares denote g-dominated $l=1$ modes; an inverted triangle marks the most p-dominated $l=1$ mode in each radial order.}
\label{f6}
\end{figure*}

The frequency \'{e}chelle diagrams in Figure~\ref{f6} show that the representative models reproduce the g-dominated mixed-mode pattern, while the radial, quadrupole, and p-dominated modes exhibit larger systematic offsets. We assess these offsets with the inverse-plus-cubic surface correction of \citet{2014A&A...568A.123B},
\begin{equation}
\delta\nu=\left (a_{-1}(\nu/\nu_{ac})^{-1}
+a_{3}(\nu/\nu_{ac})^3 \right )/I,
\label{e13}
\end{equation}

where $a_{-1}$ and $a_3$ are fitted coefficients, $I$ is the normalized inertia from equation~(\ref{e2}), and the acoustic cutoff frequency is scaled as
\begin{equation}
\nu_{ac}/\nu_{ac,\odot}=\frac{g}{g_{\odot}}\left ( 
\frac{T_{\rm eff}}{T_{\rm eff,\odot}}\right )^{-1/2},
\label{e14}
\end{equation} 

We adopt $\nu_{ac,\odot}=5000\,\mu$Hz \citep{2014A&A...568A.123B}, $\log g_\odot=4.44$, and $T_{\rm eff,\odot}=5777$~K \citep{2000asqu.book.....C}. The corrections alter the g-dominated frequencies negligibly but are larger for the acoustic and p-dominated modes. Thus, within the present precision, $\chi^2_{l=1g}$ can be evaluated without applying a surface correction.

\begin{figure*}[t]
\centering
\includegraphics[angle=0,scale=1.0]{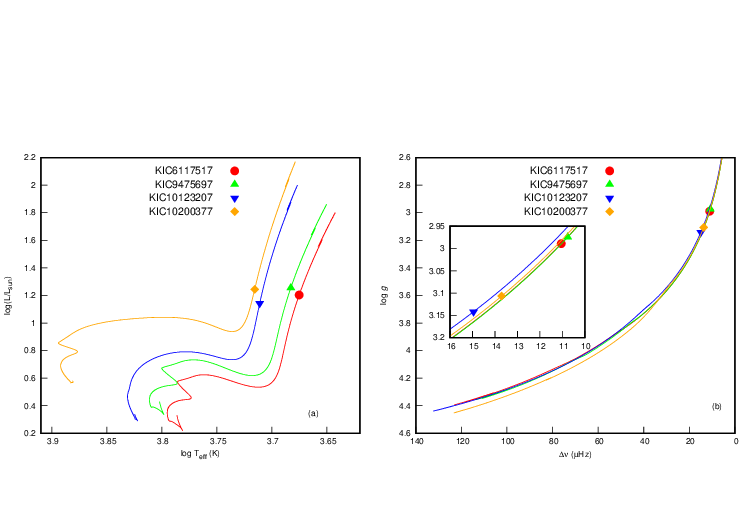}
\caption{Evolutionary tracks of KIC~6117517, KIC~9475697, KIC~10123207, and KIC~10200377 in the HR diagram (panel a) and in the $\Delta\nu$--$\log g$ diagram (panel b). Colors and symbols correspond to the four targets as indicated in the legends. Filled symbols mark the positions of the final best-fitting models (Models~A, B, E, and F in Table~\ref{t7}). The tracks in panel~(a) are extended beyond the RGB bump. The inset in panel~(b) enlarges the region containing the four best-fitting models and their corresponding evolutionary tracks.}
\label{f7}
\end{figure*}

\begin{figure}[ht]
\centering
\includegraphics[angle=0,scale=1.0]{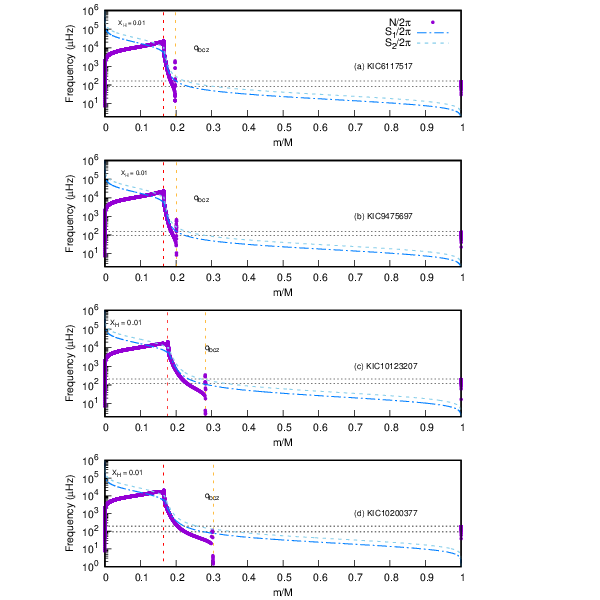}
\caption{Propagation diagrams of the best-fitting models for KIC~6117517 (panel a), KIC~9475697 (panel b), KIC~10123207 (panel c), and KIC~10200377 (panel d). The vertical red dashed lines denote the helium core boundary where the hydrogen mass fraction is about 0.01. The vertical orange dashed lines denote the base of the convection zone, marked as $q_{\rm bcz}$. Two horizontal dashed lines represent the range of observed frequencies. The purple dots indicate the Brunt-V\"{a}is\"{a}l\"{a} frequency $N$, and the blue dash-dotted and light-blue dashed lines indicate the characteristic acoustic 
frequencies $S_{1}$ and $S_{2}$, respectively.}
\label{f8}
\end{figure}

Figure~\ref{f7} shows the evolutionary tracks of the four target stars in the HR diagram (panel~a) and in the $\Delta\nu$--$\log g$ diagram (panel~b), with their best-fitting model positions marked by filled symbols. The extended tracks show that all four best-fitting models lie on the ascending red-giant branch before the RGB bump. Their locations near the base of the RGB, with $T_{\rm eff} \approx 4734$--$5196$~K and $\log(L/L_\odot) \approx 1.14$--$1.25$, are consistent with the hydrogen shell-burning phase. Their similar positions in the $\Delta\nu$--$\log g$ diagram indicate that they are at broadly similar evolutionary stages, as further supported by their nearly identical helium-core masses ($M_{\rm He}\approx0.197$--$0.210\,M_\odot$; Table~\ref{t7}). Figure~\ref{f8} presents the propagation diagrams of the four best-fitting models. A notable structural difference among the four models is the location of the convection-zone base $q_{\rm bcz}$ (defined as the mass coordinate where $\nabla_r = \nabla_a$): it lies closer to the helium-core boundary in KIC~6117517 and KIC~9475697 than in KIC~10123207 and KIC~10200377, consistent with their more advanced evolutionary states as reflected by their smaller large frequency separations ($\Delta\nu = 10.078$ and $9.786~\mu$Hz, respectively) compared to those of KIC~10123207 and KIC~10200377 ($\Delta\nu = 13.566$ and $12.406~\mu$Hz, respectively). The detailed internal structure profiles are presented in Figures~\ref{fa1}--\ref{fa4}.

\begin{table*}
\caption{\label{t7}Input and fundamental parameters of the final best-fitting models. Models A and B are from Grid~A; Models E and F are from Grid~B (see Table~\ref{t5} for Models C and D from Grid~A).}
\centering
\begin{tabular}{lcccc}
\hline
\hline
Parameters                            &KIC6117517  &KIC9475697 &KIC10123207 &KIC10200377 \\
                                      &Model A    &Model B    &Model E     &Model F  \\
\hline
&&Input Parameters\\
\hline
Mass (M/$M_{\odot}$)                  &1.26       &1.28       &1.12         &1.25\\
Initial Metal fraction (Z)            &0.025      &0.017      &0.005        &0.003\\
Overshooting parameter ($f_{ov}$)     &0.007      &0.006      &0.006        &0.006 \\
Mixing length parameter ($\alpha_{\rm MLT}$)    &2.0        &2.0        &2.2          &2.2\\
\hline
&&$\chi^{2}$ Value\\
\hline
$\chi_{l=1g}^{2}$             &1392.159   &555.807    &3756.599  &7582.781\\
$\chi_{all}^{2}$              &1092.425   &558.047    &2336.612  &5130.774\\
$S^{2}_{l=1g}$                &0.00138    &0.00198    &0.000744  &0.000831\\
$S^{2}_{all}$                 &0.0382     &0.0167     &0.00632   &0.0481\\
\hline
&&Fundamental Parameters\\
\hline
Effective temperature ($\rm T_{eff}$/K)              
                              &4733.6    &4817.1    &5141.0    &5195.7 \\
$\rm [Fe/H]_{mod}$            &0.224     &0.052     &-0.441    &-0.647\\
Surface gravity $\log g$ (dex)   &2.989     &2.975     &3.142     &3.107\\
Radius (R/$R_{\odot}$)        &5.951     &6.099     &4.706     &5.177 \\
Luminosity ($\rm L_{\odot}$)  &15.98     &18.00     &13.90     &17.55\\
Large frequency separation ($\Delta\nu_{m}$/$\mu$Hz) 
                              &10.078    &9.786     &13.566    &12.406\\
Period spacing ($\Delta\Pi_{1,mod}$/s)               
                              &76.932    &75.702    &83.757    &81.509\\
Age (t/Gyr)                   &5.650     &4.717     &5.185     &3.179 \\
Mass of helium core ($M_{He}$/$M_{\odot}$)           
                              &0.2071    &0.2104    &0.1971    &0.2053 \\
Radius of helium core ($R_{He}$/$R_{\odot}$)         
                              &0.03113   &0.03073   &0.03121   &0.03120\\
\hline
\end{tabular}
\end{table*}

\subsection{Convective Core Overshooting}\label{sec4.4}

\begin{figure*}[t]
\centering
\includegraphics[angle=0,width=\textwidth]{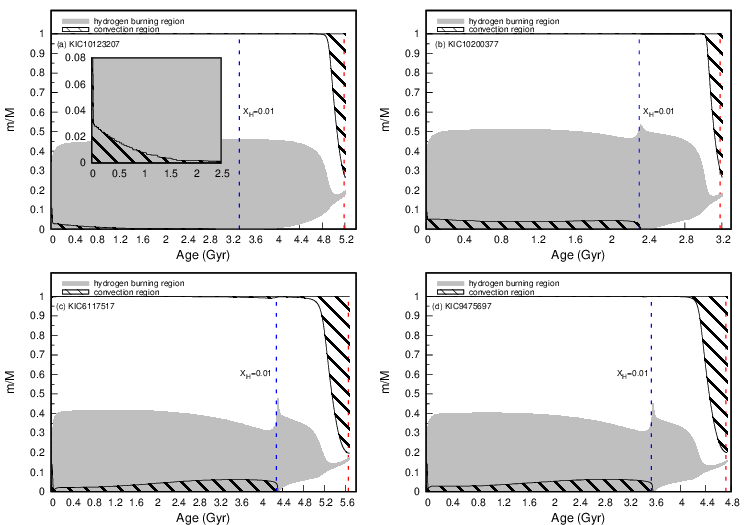}
\caption{Kippenhahn diagrams of the best-fitting models for KIC~10123207, KIC~10200377, KIC~6117517, KIC~9475697, ordered by increasing stellar mass (panels a--d). Hatched regions indicate convection zones; grey shading indicates hydrogen-burning regions (nuclear energy generation $> 1~{\rm erg\,g^{-1}\,s^{-1}}$). Vertical blue dashed lines mark the end of the main sequence ($X_{\rm H} = 0.01$); vertical red dashed lines indicate the current age of each star.}
\label{f9}
\end{figure*}

Figure~\ref{f9} presents the Kippenhahn diagrams of the four best-fitting models, ordered by increasing stellar mass, illustrating the temporal evolution of convective regions from the main sequence to the RGB. The extent of the main-sequence convective core increases monotonically with mass; the inset of panel~(a) highlights the particularly small 
convective core of KIC~10123207, the least massive star in our sample ($1.12\,M_\odot$).

The best-fitting overshooting parameters for the four new targets (Table~\ref{t7}) are $f_{\rm ov} = 0.006$ for KIC~10123207 and KIC~10200377, $f_{\rm ov} = 0.007$ for KIC~6117517, and $f_{\rm ov} = 0.006$ for KIC~9475697. Combined with the four red giants from Paper~I---KIC~9970396 ($1.14\,M_\odot$, $f_{\rm ov} = 0.005$), KIC~9145955 ($1.20\,M_\odot$, $f_{\rm ov} = 0.007$), KIC~11968334 ($1.31\,M_\odot$, $f_{\rm ov} = 0.009$), and KIC~9882316 ($1.40\,M_\odot$, $f_{\rm ov} = 0.015$)---our sample now comprises eight red giants spanning $1.12$--$1.40\,M_\odot$.

\begin{figure}[ht]
\centering
\includegraphics[angle=0,width=0.48\textwidth]{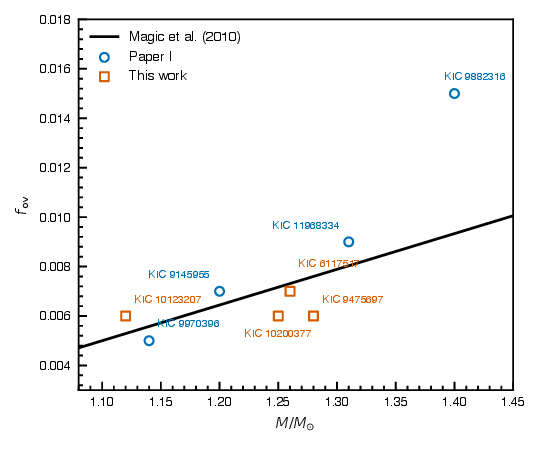}
\caption{Overshooting parameter $f_{\rm ov}$ as a function of stellar mass $M$ for the eight red giants analyzed in 
Paper~I (blue circles) and this work (red squares). The solid line represents the linear mass dependence proposed by \citet{2010ApJ...718.1378M}, $f_{\rm ov} = (0.13\,M/M_{\odot} - 0.098)/9$, derived from stellar evolution modeling of open clusters. The results from both papers are in broad agreement with the general trend of increasing overshooting 
efficiency with stellar mass.}
\label{f10}
\end{figure}

Figure~\ref{f10} shows $f_{\rm ov}$ as a function of stellar mass for all eight stars, together with the theoretical linear relation of \citet{2010ApJ...718.1378M}: $f_{\rm ov} = (0.13\,M/M_\odot - 0.098)/9$, derived from open-cluster modeling and applicable over $1.1$--$2.0\,M_\odot$. Across our mass range, this relation predicts $f_{\rm ov} \approx 
0.005$--$0.009$; seven of our eight values fall within this range. The best-fitting overshooting parameters for the four new targets ($f_{\rm ov} = 0.006$--$0.007$; Table~\ref{t7}) are consistent with the mass-dependent upper bound of $f_{\rm ov} \lesssim 0.010$ established for stars with $M \lesssim 1.4\,M_\odot$ in Paper~I, within whose grid bounds the present search was conducted. The exception among the combined sample is KIC~9882316 ($1.40\,M_\odot$, $f_{\rm ov} = 0.015$), which exceeds both the \citet{2010ApJ...718.1378M} prediction and the typical values found for less massive stars, and may reflect enhanced or variable overshooting efficiency near this mass threshold. Taken together, the combined sample shows a positive $f_{\rm ov}$--mass trend consistent with the prediction of \citet{2010ApJ...718.1378M} and with independent results from double-lined eclipsing binary analyses \citep{2017ApJ...849...18C, 2018ApJ...859..100C, 2019ApJ...876..134C}, supporting a mass-dependent overshooting efficiency in the range $1.1$--$1.4\,M_\odot$.

\subsection{Internal Rotation of KIC~9475697}\label{sec4.5}

Among the four targets, KIC~9475697 is the only star for which sufficient rotational multiplets are detected to permit a rotation analysis; the measured splittings are listed in Table~\ref{ta2}. For slowly rotating evolved stars, second-order effects are negligible, and the first-order relation can be expressed as \citep{2010aste.book.....A}
\begin{equation}
\nu_{nlm} = \nu_{nl0} + m\,\delta\nu_{nl},
\label{e15}
\end{equation}
where $\delta\nu_{n,l}$ is the rotational splitting. \citet{2013A&A...549A..75G} showed that the rotational splitting depends linearly on the mode-trapping parameter $\zeta \equiv I_g/I$:
\begin{equation}
\delta\nu_{nl} = \zeta\!\left(\frac{\Omega_{\rm core}}{2}
               - \Omega_{\rm env}\right) + \Omega_{\rm env},
\label{e16}
\end{equation}
where $\Omega_{\rm core}$ and $\Omega_{\rm env}$ are the mean cyclic rotation frequencies of the radiative core and convective envelope, respectively. Modes with $\zeta \approx 1$ predominantly probe the core, while those with smaller $\zeta$ are more sensitive to the envelope \citep{2012A&A...548A..10M}.

We estimate $\zeta$ by two independent methods that serve as a mutual consistency check. The first, $\zeta_{\rm mod}$, is computed from the displacement eigenfunctions of the best-fitting stellar model (Model~B; Table~\ref{t7}):
\begin{equation}
\zeta_{\rm mod} \equiv \frac{I_{g}}{I} =
\frac{\displaystyle\int_{r_1}^{r_2}
  \bigl[\xi_{r}^{2}+l(l+1)\xi_{h}^{2}\bigr]
  \rho r^{2}\,dr}
{\displaystyle\int_{0}^{R}
  \bigl(\xi_{r}^{2}+l(l+1)\xi_{h}^{2}\bigr)
  \rho r^{2}\,dr},
\label{e17}
\end{equation}
where $r_1$ and $r_2$ are the boundaries of the g-mode cavity, $\xi_r$ and $\xi_h$ are the radial and horizontal displacement eigenfunctions, $\rho$ is the local density, and $R$ is the stellar radius. Because the p-mode component of a mixed-mode eigenfunction is sensitive to near-surface structural uncertainties \citep[the surface 
effect;][]{Ong2024}, $\zeta_{\rm mod}$ may systematically bias the inferred envelope rotation rate. The second estimate, $\zeta_{\rm as}$, is derived entirely from observed frequencies and is therefore free of this bias 
\citep{2013A&A...549A..75G,2014A&A...564A..27D,2015A&A...584A..50M}:

\begin{equation}
\zeta_{\rm as} = \frac{\Delta P_{\rm obs}}
                      {\Delta\Pi_{1,\rm obs}},
\label{e18}
\end{equation}
where $\Delta P_{\rm obs}$ is the observed period spacing between consecutive $l=1$ mixed modes and $\Delta\Pi_{1,\rm obs}$ is the asymptotic dipole period spacing. 

We fit equation~(\ref{e16}) in the form $\delta\nu = a\zeta + b$, from which $\Omega_{\rm env} = b$ and 
$\Omega_{\rm core} = 2(a+b)$, weighted by the measured splitting uncertainties. Both methods start from the same 19 observed multiplets. For $\zeta_{\rm mod}$, the initial fit yields a negative $\Omega_{\rm env}$; we therefore remove modes sequentially by decreasing standardized residual, stopping when $\Omega_{\rm env} \geq 0$, which removes four outliers and leaves 15 modes. For $\zeta_{\rm as}$, we form 14 adjacent-mode pairs, evaluate $\zeta_{\rm as}$ at the harmonic-mean frequency using $\Delta\Pi_{1,\rm obs} = 75.70$~s, exclude two pairs with $\zeta_{\rm as} > 1$ as unphysical, and apply the same residual criterion to the remaining 12 pairs, leaving 10. The retained multiplets and pairs are listed in Tables~\ref{t8} and \ref{t9}. Note that because $\zeta_{\rm mod}$ operates on individual modes while $\zeta_{\rm as}$ operates on adjacent-mode pairs, the two retained subsets do not exactly correspond.

\begin{figure}[ht]
\centering
\includegraphics[angle=0,scale=1.0]{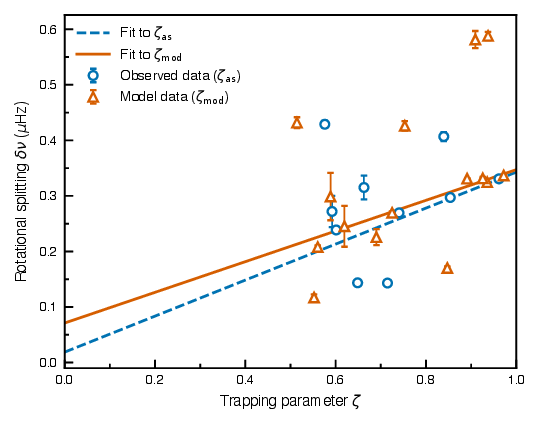}
\caption{Rotational splittings $\delta\nu_{nl}$ as a function of the mode-trapping parameter $\zeta$ for KIC~9475697. Blue open circles and the dashed blue line show the splittings and weighted linear fit using the observation-based $\zeta_{\rm as}$; orange open triangles and the solid orange line show the corresponding results using the model-derived $\zeta_{\rm mod}$. }
\label{f11}
\end{figure}

Figure~\ref{f11} shows the rotational splittings $\delta\nu_{nl}$ as a function of the trapping parameter $\zeta$ for both 
methods, together with the weighted linear fits. Both fits exhibit a clear positive slope, confirming that g-dominated modes 
($\zeta \to 1$) carry systematically larger splittings than p-dominated modes, consistent with the theoretical expectation of a rapidly rotating core. The numerical results are listed in Table~\ref{t10}: the two methods yield core rotation rates of 
$\Omega_{\rm core} = 0.6945\pm0.0190\,\mu$Hz ($\zeta_{\rm mod}$) and $0.6857\pm0.0193\,\mu$Hz ($\zeta_{\rm as}$), differing by only $\sim\!1.3\%$. This mutual consistency, despite the two methods relying on entirely different inputs, namely model-derived mode inertias and observed period spacings, lends support to the reliability of our stellar model and indicates that the g-dominated mixed-mode frequencies are well reproduced by the best-fitting model. We therefore adopt the model-based estimate, $\Omega_{\rm core} = 0.6945\pm0.0190\,\mu$Hz, as our primary result.

The corresponding envelope rates, $0.071 \pm 0.094\,\mu$Hz and $0.019 \pm 0.077\,\mu$Hz, carry large uncertainties, as expected given that the predominantly g-dominated modes carry little sensitivity to the convective envelope \citep{2012ApJ...756...19D,2013A&A...549A..75G,2021RvMP...93a5001A}. The offset between the two $\Omega_{\rm env}$ central values may partly reflect a surface-effect bias in $\zeta_{\rm mod}$: near-surface structural deficiencies alter the p-mode contribution to the mixed-mode eigenfunctions and preferentially bias the low-$\zeta$ end of the fit where $\Omega_{\rm env}$ is determined \citep{Ong2024}. In contrast, $\zeta_{\rm as}$ avoids direct dependence on model frequencies, but remains sensitive to missing mixed modes and departures from the asymptotic approximation. The consistency between the two envelope estimates, despite these complementary limitations, provides confidence that no large systematic error affects either result.

Table~\ref{t10} compares our results with those of \citet{LiG2024}, who measured rotation rates for more than 2000 
red giants using a global asymptotic trapping parameter $\zeta_\nu$ constructed from fitted mixed-mode pattern parameters ($\Delta\Pi_1$, $\Delta\nu$, $q$, $\theta_p$), obtaining $\Omega_{\rm core} = 1.287\pm0.009\,\mu$Hz and $\Omega_{\rm env} = -0.015\pm0.013\,\mu$Hz for KIC~9475697. Their core rate is $\sim\!85\%$ higher than our model-based value. Our analysis differs in two key respects: first, $\zeta_{\rm mod}$ is derived from a stellar model that reproduces the individual mixed-mode frequencies of KIC~9475697 to high precision (Section~\ref{sec4.3}), making it more precise than a global asymptotic fit; second, the independent $\zeta_{\rm as}$ estimate, which relies solely on observed frequencies, yields a result consistent with $\zeta_{\rm mod}$, further supporting our lower core rotation rate. The $\sim\!85\%$ discrepancy may arise from differences in the definition and computation of $\zeta$, mode selection, and fitting methodology. Near-degeneracy effects \citep[e.g.,][]{2015A&A...580A..96D, 2017A&A...600A...1M, 2017A&A...605A..75D} may additionally affect individual splitting measurements.

\begin{deluxetable*}{cccccc}
\tablecaption{Dipole rotational multiplets used in 
the $\zeta_{\rm mod}$ fit for KIC~9475697\label{t8}}
\tablehead{
  \colhead{Multiplet} &
  \colhead{$\nu_{m=-1}$} &
  \colhead{$\nu_{m=0}$} &
  \colhead{$\nu_{m=+1}$} &
  \colhead{$\delta\nu$} &
  \colhead{$\zeta_{\rm mod}$} \\
  \colhead{} &
  \multicolumn{3}{c}{($\mu$Hz)} &
  \colhead{($\mu$Hz)} &
  \colhead{}
}
\startdata
1  & 105.6149 & 105.7226 & 105.8487 & 0.1169 & 0.5522 \\
2  & 106.1042 & 106.2743 & \nodata  & 0.1701 & 0.8473 \\
3  & 113.0133 & 113.3324 & 113.6867 & 0.3367 & 0.9721 \\
4  & 113.9444 & 114.2752 & 114.5938 & 0.3247 & 0.9356 \\
5  & 114.8541 & 115.1256 & 115.3931 & 0.2695 & 0.7251 \\
6  & 115.5532 & 115.7320 & 115.9689 & 0.2079 & 0.5604 \\
7  & 116.1566 & 116.4876 & \nodata  & 0.3310 & 0.8912 \\
8  & 124.4225 & 124.9012 & 125.2748 & 0.4262 & 0.7531 \\
9  & \nodata  & 125.5852 & 126.0170 & 0.4318 & 0.5143 \\
10 & 133.3517 & 133.9354 & 134.5278 & 0.5881 & 0.9376 \\
11 & 134.9238 & 135.0852 & 135.3753 & 0.2258 & 0.6902 \\
12 & 136.3724 & 136.9539 & \nodata  & 0.5815 & 0.9093 \\
13 & \nodata  & 144.1649 & 144.4964 & 0.3315 & 0.9258 \\
14 & \nodata  & 145.2152 & 145.5140 & 0.2988 & 0.5888 \\
15 & \nodata  & 146.1664 & 146.4116 & 0.2452 & 0.6195
\enddata
\tablecomments{All modes have $l=1$; the azimuthal order $m$ is indicated by the column heading. A missing component is denoted by \nodata. For complete triplets, $\delta\nu = (\nu_{m=+1}-\nu_{m=-1})/2$; for doublets, $\delta\nu$ is the separation between the detected side lobe and the central component, with the sign chosen so that $\delta\nu > 0$. Four modes were removed prior to this fit by the sequential residual criterion described in Section~\ref{sec4.5}; they are centered at $\nu_{m=0} = 121.698$, $117.454$, $107.034$, and $105.121\,\mu$Hz. The formal uncertainties from this filtered fit do not account for mode-selection uncertainty.}
\end{deluxetable*}

\begin{deluxetable*}{cccccc}
\tablecaption{Observed dipole-mode pairs used in 
the $\zeta_{\rm as}$ fit for KIC~9475697\label{t9}}
\tablehead{
  \colhead{Pair} &
  \colhead{$\nu_{i}$} &
  \colhead{$\nu_{i+1}$} &
  \colhead{$\nu_{\Delta P}$} &
  \colhead{$\zeta_{\rm as}$} &
  \colhead{$\overline{\delta\nu}$} \\
  \colhead{} &
  \multicolumn{3}{c}{($\mu$Hz)} &
  \colhead{} &
  \colhead{($\mu$Hz)}
}
\startdata
1  & 105.1212 & 105.7226 & 105.4210 & 0.7148 & 0.1431 \\
2  & 105.7226 & 106.2743 & 105.9977 & 0.6487 & 0.1435 \\
3  & 113.3324 & 114.2752 & 113.8018 & 0.9617 & 0.3307 \\
4  & 114.2752 & 115.1256 & 114.6988 & 0.8539 & 0.2971 \\
5  & 115.1256 & 115.7320 & 115.4280 & 0.6012 & 0.2387 \\
6  & 115.7320 & 116.4876 & 116.1086 & 0.7404 & 0.2694 \\
7  & 124.9012 & 125.5852 & 125.2423 & 0.5760 & 0.4290 \\
8  & 133.9354 & 135.0852 & 134.5078 & 0.8395 & 0.4069 \\
9  & 144.1649 & 145.2152 & 144.6881 & 0.6627 & 0.3152 \\
10 & 145.2152 & 146.1664 & 145.6892 & 0.5920 & 0.2720
\enddata
\tablecomments{Both frequencies in each pair are observed $l=1$, $m=0$ modes. The trapping parameter $\zeta_{\rm as}$ is evaluated at the harmonic-mean frequency $\nu_{\Delta P}=2(\nu_i^{-1}+\nu_{i+1}^{-1})^{-1}$, using $\Delta\Pi_{1,\rm obs}=75.70$~s, and $\overline{\delta\nu}$ is the mean rotational splitting of the two adjacent multiplets. Of the original 14 pairs, two at $121.698$--$124.901$ and $135.085$--$136.954\,\mu$Hz were excluded because they yield $\zeta_{\rm as}>1$; two further pairs at $116.488$--$117.454$ and $106.274$--$107.034\,\mu$Hz were removed by the sequential residual criterion. }
\end{deluxetable*}

\begin{table}[ht!]
\caption{Internal Rotation Rates of 
KIC~9475697\label{t10}}
\centering
\setlength{\tabcolsep}{8pt}
\renewcommand{\arraystretch}{1.25}
\begin{tabular}{lccc}
\hline\hline
 & & \multicolumn{2}{c}{Rotation Rate ($\mu$Hz)} \\
\cmidrule(lr){3-4}
Parameter & Method & This Work & \citet{LiG2024} \\
\midrule
$\Omega_{\rm core}$ & $\zeta_{\rm mod}$ 
    & $0.695 \pm 0.019$ & \nodata \\
$\Omega_{\rm core}$ & $\zeta_{\rm as}$  
    & $0.686 \pm 0.019$ & \nodata \\
$\Omega_{\rm core}$ & $\zeta_{\nu}$     
    & \nodata           & $1.287 \pm 0.009$ \\
\midrule
$\Omega_{\rm env}$  & $\zeta_{\rm mod}$ 
    & $0.071 \pm 0.094$ & \nodata \\
$\Omega_{\rm env}$  & $\zeta_{\rm as}$  
    & $0.019 \pm 0.077$ & \nodata \\
$\Omega_{\rm env}$  & $\zeta_{\nu}$     
    & \nodata           & $-0.015 \pm 0.013$ \\
\bottomrule
\end{tabular}
\tablecomments{$\zeta_{\rm mod}$ is computed from the displacement eigenfunctions of the best-fitting stellar model (Model~B; Table~\ref{t7}); $\zeta_{\rm as}$ is derived from consecutive observed period spacings and is independent of model assumptions; $\zeta_{\nu}$ \citep{LiG2024} is obtained from a global asymptotic fit to the observed mixed-mode pattern using $\Delta\Pi_1$, 
$\Delta\nu$, $q$, and $\theta_p$. }
\end{table}

\section{Conclusions}\label{sec5}

Based on the convective mixing and overshooting constraints from Paper~I, we modeled four lower-mass \textit{Kepler} red giants spanning a broad metallicity range using $l=1$ g-dominated mixed modes, aiming to test whether those constraints extend to a broader metallicity range, to constrain their global and internal stellar properties, and to probe the core rotation of KIC~9475697. Our principal findings are as follows:

\begin{enumerate}

\item Grid~A with fixed $\alpha_{\rm MLT}=2.0$ reproduces the observations of the metal-rich targets KIC~6117517 ($\rm[Fe/H]=+0.2224$) and KIC~9475697 ($\rm[Fe/H]=+0.0456$), suggesting that the value calibrated in Paper~I remains appropriate for this metallicity regime. For the two metal-poor targets KIC~10123207 ($\rm[Fe/H]=-0.45$) and KIC~10200377 ($\rm[Fe/H]=-0.5812$), Grid~B (spectroscopic metallicity fixed, $\alpha_{\rm MLT}$ free) yields $\alpha_{\rm MLT}\approx 2.16$--$2.18$. The resulting trend, namely that lower metallicity requires higher $\alpha_{\rm MLT}$, is qualitatively consistent with predictions from 3D hydrodynamical simulations \citep{2014MNRAS.445.4366T,2015A&A...573A..89M} but opposite to some 1D empirical calibrations \citep[e.g.,][]{2024ApJ...974...77L}. Given our limited sample of four stars, this result should be treated as suggestive; a larger, systematic analysis will be needed to establish the statistical significance of the trend.

\item The grid modeling yields well-constrained fundamental and helium-core parameters. Conditional on the adopted input physics and grid sampling, the likelihood-weighted results from Grid~A (Table~\ref{t4}) reach formal relative uncertainties of approximately $0.6$--$2.5\%$ in mass, $0.2$--$0.9\%$ in radius, and $2$--$13\%$ in age. The helium-core masses span $M_{\rm He}\approx0.195$--$0.210\,M_\odot$, with formal relative uncertainties below $0.6\%$, while the helium-core radii span $R_{\rm He}\approx0.0307$--$0.0312\,R_\odot$, with formal relative uncertainties below $0.2\%$. 

\item Surface-effect corrections have a negligible impact on g-dominated mixed modes. The application of the \citet{2014A&A...568A.123B} surface correction results in negligible changes to the g-dominated mode frequencies, confirming that 
$\chi^2_{l=1g}$ provides a robust and computationally efficient metric for constraining stellar models without requiring surface-term corrections.

\item The core overshooting parameters are consistent with a mass-dependent overshooting relation. The best-fitting values range from $f_{\rm ov}=0.006$ to $0.007$, in reasonable agreement with the linear mass--$f_{\rm ov}$ relation of \citet{2010ApJ...718.1378M}, and all lie within the constraint $f_{\rm ov}<0.01$ derived for stars with $M<1.4\,M_{\odot}$ in Paper~I. The small residual discrepancies are attributable to the finite grid resolution and to the well-known difficulty of constraining overshooting precisely from individual stars.

\item The rotational multiplets of KIC~9475697 yield mutually consistent estimates of the near-core rotation rate from two independent trapping parameters. The model-based and observation-based analyses give $\Omega_{\rm core}=0.6945\pm0.0190\,\mu$Hz 
and $0.6857\pm0.0193\,\mu$Hz, respectively, agreeing within $1\sigma$ ($\sim\!1.3\%$). We adopt the model-based estimate as our 
primary result. 

\end{enumerate}

The present results demonstrate that the $\chi^{2}_{l=1g}$ framework developed in Paper~I can be applied to low-luminosity RGB stars over the metallicity range sampled here. Future work could extend this framework to larger samples with well-resolved dipole mixed-mode spectra and reliable spectroscopic constraints. The \textit{Kepler} archive provides the most immediate opportunity for extending the present analysis. Large samples of RGB stars have already been characterized using radial-mode frequencies, gravity-mode period spacings, Gaia luminosities, and spectroscopic constraints \citep{2023ApJ...953..182W}, while individual mixed-mode properties have been measured for more than 1000 low-luminosity red giants \citep{2023ApJ...954..152K}. These catalogs provide a well-defined pool from which a high-quality subset can be selected for detailed individual-frequency modeling across broader ranges of mass and metallicity. Such an analysis would enable statistical tests of the tentative $\alpha_{\rm MLT}$--metallicity trend found here and of the suggested mass dependence of $f_{\rm ov}$.

Using \textit{TESS} observations, \citet{2025ApJS..279...37Z} extracted individual frequencies of radial, dipole, and quadrupole modes and derived asymptotic mixed-mode parameters for 687 red giants. Their results demonstrate that \textit{TESS} can provide the dipole mixed-mode information needed to extend the present $\chi^{2}_{l=1g}$ framework beyond the \textit{Kepler} field to suitable low-luminosity RGB stars observed by \textit{TESS}. The long-duration, high-precision photometry expected from \textit{PLATO} should be particularly well suited to extending the present individual-frequency modeling method to a substantially larger sample \citep{2025ExA....59...26R}.

\textit{Roman} is expected to extend red-giant asteroseismology to the distant and crowded Galactic bulge and to provide global asteroseismic parameters for large stellar samples \citep{2025ApJ...987..181W}. However, its nominal seasonal observing windows are not expected to support the detailed ``boutique'' asteroseismology required to probe internal structure and rotation \citep{2023arXiv230703237H}. \textit{Roman} may therefore provide population-level constraints and identify targets for follow-up, whereas direct application of the present individual-frequency method would require longer, more nearly continuous time series than those provided by the nominal survey.

\begin{acknowledgments}
This work is supported by the National Key R\&D Program of China No. 2024YFA1611901, NSFC grants (12090040, 12090042, 12603051), the Joint Research Fund in Astronomy (U2031203) under cooperative agreement between the National Natural Science Foundation of China (NSFC) and the Chinese Academy of Sciences (CAS), the Fundamental Research Funds for the Central Universities, and the CSST project (CMS-CSST-2025-A15). This work is also supported by the Postdoctoral Fellowship Program of CPSF under Grant Number GZC20240126 and the China Postdoctoral Science Foundation under Grant Number 2024M760241. The authors acknowledge NASA and the \textit{Kepler} team for making the \textit{Kepler} data available. The \textit{Kepler} Mission is funded by NASA’s Science Mission Directorate.

The MESA r10398 work directories, input inlists, project-specific
\texttt{run\_star\_extras.f} files, and the MESA evolutionary-track and
ADIPLS oscillation output files retained for Models~A--F have been deposited
in the AAS Journals Zenodo community and will become publicly available
following curator approval:
\dataset[doi:10.5281/zenodo.23310560]{https://doi.org/10.5281/zenodo.23310560}.
\end{acknowledgments}



\bibliography{zxyref}{}

@ARTICLE{2022ApJS..259...35A,
       author = {{Abdurro'uf} and {Accetta}, Katherine and {Aerts}, Conny and {Silva Aguirre}, V{\'\i}ctor and {Ahumada}, Romina and {Ajgaonkar}, Nikhil and {Filiz Ak}, N. and {Alam}, Shadab and {Allende Prieto}, Carlos and {Almeida}, Andr{\'e}s and et al.},
        title = "{The Seventeenth Data Release of the Sloan Digital Sky Surveys: Complete Release of MaNGA, MaStar, and APOGEE-2 Data}",
      journal = {\apjs},
         year = 2022,
        month = apr,
       volume = {259},
       number = {2},
          eid = {35},
        pages = {35},
          doi = {10.3847/1538-4365/ac4414},
archivePrefix = {arXiv},
       eprint = {2112.02026},
 primaryClass = {astro-ph.GA},
       adsurl = {https://ui.adsabs.harvard.edu/abs/2022ApJS..259...35A}
}

@BOOK{2010aste.book.....A,
       author = {{Aerts}, Conny and {Christensen-Dalsgaard}, J{\o}rgen and {Kurtz}, Donald W.},
        title = "{Asteroseismology}",
         year = 2010,
          doi = {10.1007/978-1-4020-5803-5},
       adsurl = {https://ui.adsabs.harvard.edu/abs/2010aste.book.....A}
}

@ARTICLE{2021RvMP...93a5001A,
       author = {{Aerts}, C.},
        title = "{Probing the interior physics of stars through asteroseismology}",
      journal = {Reviews of Modern Physics},
         year = 2021,
        month = jan,
       volume = {93},
       number = {1},
          eid = {015001},
        pages = {015001},
          doi = {10.1103/RevModPhys.93.015001},
archivePrefix = {arXiv},
       eprint = {1912.12300},
 primaryClass = {astro-ph.SR},
       adsurl = {https://ui.adsabs.harvard.edu/abs/2021RvMP...93a5001A}
}

@ARTICLE{2014A&A...568A.123B,
       author = {{Ball}, W.~H. and {Gizon}, L.},
        title = "{A new correction of stellar oscillation frequencies for near-surface effects}",
      journal = {\aap},
         year = 2014,
        month = aug,
       volume = {568},
          eid = {A123},
        pages = {A123},
          doi = {10.1051/0004-6361/201424325},
archivePrefix = {arXiv},
       eprint = {1408.0986},
 primaryClass = {astro-ph.SR},
       adsurl = {https://ui.adsabs.harvard.edu/abs/2014A&A...568A.123B}
}

@ARTICLE{2012Natur.481...55B,
       author = {{Beck}, Paul G. and {Montalban}, Josefina and {Kallinger}, Thomas and {De Ridder}, Joris and {Aerts}, Conny and {Garc{\'\i}a}, Rafael A. and {Hekker}, Saskia and {Dupret}, Marc-Antoine and {Mosser}, Benoit and {Eggenberger}, Patrick and {Stello}, Dennis and {Elsworth}, Yvonne and {Frandsen}, S{\o}ren and {Carrier}, Fabien and {Hillen}, Michel and {Gruberbauer}, Michael and {Christensen-Dalsgaard}, J{\o}rgen and {Miglio}, Andrea and {Valentini}, Marica and {Bedding}, Timothy R. and {Kjeldsen}, Hans and {Girouard}, Forrest R. and {Hall}, Jennifer R. and {Ibrahim}, Khadeejah A.},
        title = "{Fast core rotation in red-giant stars as revealed by gravity-dominated mixed modes}",
      journal = {\nat},
         year = 2012,
        month = jan,
       volume = {481},
       number = {7379},
        pages = {55-57},
          doi = {10.1038/nature10612},
archivePrefix = {arXiv},
       eprint = {1112.2825},
 primaryClass = {astro-ph.SR},
       adsurl = {https://ui.adsabs.harvard.edu/abs/2012Natur.481...55B}
}

@ARTICLE{2011Natur.471..608B,
       author = {{Bedding}, Timothy R. and {Mosser}, Benoit and {Huber}, Daniel and {Montalb{\'a}n}, Josefina and {Beck}, Paul and {Christensen-Dalsgaard}, J{\o}rgen and {Elsworth}, Yvonne P. and {Garc{\'\i}a}, Rafael A. and {Miglio}, Andrea and {Stello}, Dennis and {White}, Timothy R. and {De Ridder}, Joris and {Hekker}, Saskia and {Aerts}, Conny and {Barban}, Caroline and {Belkacem}, Kevin and {Broomhall}, Anne-Marie and {Brown}, Timothy M. and {Buzasi}, Derek L. and {Carrier}, Fabien and {Chaplin}, William J. and {di Mauro}, Maria Pia and {Dupret}, Marc-Antoine and {Frandsen}, S{\o}ren and {Gilliland}, Ronald L. and {Goupil}, Marie-Jo and {Jenkins}, Jon M. and {Kallinger}, Thomas and {Kawaler}, Steven and {Kjeldsen}, Hans and {Mathur}, Savita and {Noels}, Arlette and {Silva Aguirre}, Victor and {Ventura}, Paolo},
        title = "{Gravity modes as a way to distinguish between hydrogen- and helium-burning red giant stars}",
      journal = {\nat},
         year = 2011,
        month = mar,
       volume = {471},
       number = {7340},
        pages = {608-611},
          doi = {10.1038/nature09935},
archivePrefix = {arXiv},
       eprint = {1103.5805},
 primaryClass = {astro-ph.SR},
       adsurl = {https://ui.adsabs.harvard.edu/abs/2011Natur.471..608B}
}

@ARTICLE{1958ZA.....46..108B,
       author = {{B{\"o}hm-Vitense}, E.},
        title = "{{\"U}ber die Wasserstoffkonvektionszone in Sternen verschiedener Effektivtemperaturen und Leuchtkr{\"a}fte. Mit 5 Textabbildungen}",
      journal = {\zap},
         year = 1958,
        month = jan,
       volume = {46},
        pages = {108},
       adsurl = {https://ui.adsabs.harvard.edu/abs/1958ZA.....46..108B}
}

@ARTICLE{2013ARA&A..51..353C,
       author = {{Chaplin}, William J. and {Miglio}, Andrea},
        title = "{Asteroseismology of Solar-Type and Red-Giant Stars}",
      journal = {\araa},
         year = 2013,
        month = aug,
       volume = {51},
       number = {1},
        pages = {353-392},
          doi = {10.1146/annurev-astro-082812-140938},
archivePrefix = {arXiv},
       eprint = {1303.1957},
 primaryClass = {astro-ph.SR},
       adsurl = {https://ui.adsabs.harvard.edu/abs/2013ARA&A..51..353C}
}

@ARTICLE{2008Ap&SS.316..113C,
       author = {{Christensen-Dalsgaard}, J{\o}rgen},
        title = "{ADIPLS{\textemdash}the Aarhus adiabatic oscillation package}",
      journal = {\apss},
         year = 2008,
        month = aug,
       volume = {316},
       number = {1-4},
        pages = {113-120},
          doi = {10.1007/s10509-007-9689-z},
archivePrefix = {arXiv},
       eprint = {0710.3106},
 primaryClass = {astro-ph},
       adsurl = {https://ui.adsabs.harvard.edu/abs/2008Ap&SS.316..113C}
}

@ARTICLE{2017ApJ...849...18C,
       author = {{Claret}, Antonio and {Torres}, Guillermo},
        title = "{The Dependence of Convective Core Overshooting on Stellar Mass: A Semi-empirical Determination Using the Diffusive Approach with Two Different Element Mixtures}",
      journal = {\apj},
         year = 2017,
        month = nov,
       volume = {849},
       number = {1},
          eid = {18},
        pages = {18},
          doi = {10.3847/1538-4357/aa8770},
archivePrefix = {arXiv},
       eprint = {1710.08417},
 primaryClass = {astro-ph.SR},
       adsurl = {https://ui.adsabs.harvard.edu/abs/2017ApJ...849...18C}
}

@ARTICLE{2018ApJ...859..100C,
       author = {{Claret}, Antonio and {Torres}, Guillermo},
        title = "{The Dependence of Convective Core Overshooting on Stellar Mass: Additional Binary Systems and Improved Calibration}",
      journal = {\apj},
         year = 2018,
        month = jun,
       volume = {859},
       number = {2},
          eid = {100},
        pages = {100},
          doi = {10.3847/1538-4357/aabd35},
archivePrefix = {arXiv},
       eprint = {1804.03148},
 primaryClass = {astro-ph.SR},
       adsurl = {https://ui.adsabs.harvard.edu/abs/2018ApJ...859..100C}
}

@ARTICLE{2019ApJ...876..134C,
       author = {{Claret}, Antonio and {Torres}, Guillermo},
        title = "{The Dependence of Convective Core Overshooting on Stellar Mass: Reality Check and Additional Evidence}",
      journal = {\apj},
         year = 2019,
        month = may,
       volume = {876},
       number = {2},
          eid = {134},
        pages = {134},
          doi = {10.3847/1538-4357/ab1589},
archivePrefix = {arXiv},
       eprint = {1904.02714},
 primaryClass = {astro-ph.SR},
       adsurl = {https://ui.adsabs.harvard.edu/abs/2019ApJ...876..134C}
}

@ARTICLE{corsaro2015,
       author = {{Corsaro}, E. and {De Ridder}, J. and {Garc{\'\i}a}, R.~A.},
        title = "{Bayesian peak bagging analysis of 19 low-mass low-luminosity red giants observed with Kepler}",
      journal = {\aap},
         year = 2015,
        month = jul,
       volume = {579},
          eid = {A83},
        pages = {A83},
          doi = {10.1051/0004-6361/201525895},
archivePrefix = {arXiv},
       eprint = {1503.08821},
 primaryClass = {astro-ph.SR},
       adsurl = {https://ui.adsabs.harvard.edu/abs/2015A&A...579A..83C}
}

@BOOK{2000asqu.book.....C,
       author = {{Cox}, Arthur N.},
        title = "{Allen's astrophysical quantities}",
         year = 2000,
       adsurl = {https://ui.adsabs.harvard.edu/abs/2000asqu.book.....C}
}

@ARTICLE{2017A&A...601A..67C,
       author = {{Creevey}, O.~L. and {Metcalfe}, T.~S. and {Schultheis}, M. and {Salabert}, D. and {Bazot}, M. and {Th{\'e}venin}, F. and {Mathur}, S. and {Xu}, H. and {Garc{\'\i}a}, R.~A.},
        title = "{Characterizing solar-type stars from full-length Kepler data sets using the Asteroseismic Modeling Portal}",
      journal = {\aap},
         year = 2017,
        month = may,
       volume = {601},
          eid = {A67},
        pages = {A67},
          doi = {10.1051/0004-6361/201629496},
archivePrefix = {arXiv},
       eprint = {1612.08990},
 primaryClass = {astro-ph.SR},
       adsurl = {https://ui.adsabs.harvard.edu/abs/2017A&A...601A..67C}
}

@ARTICLE{2009Natur.459..398D,
       author = {{De Ridder}, Joris and {Barban}, Caroline and {Baudin}, Fr{\'e}d{\'e}ric and {Carrier}, Fabien and {Hatzes}, Artie P. and {Hekker}, Saskia and {Kallinger}, Thomas and {Weiss}, Werner W. and {Baglin}, Annie and {Auvergne}, Michel and {Samadi}, R{\'e}za and {Barge}, Pierre and {Deleuil}, Magali},
        title = "{Non-radial oscillation modes with long lifetimes in giant stars}",
      journal = {\nat},
         year = 2009,
        month = may,
       volume = {459},
       number = {7245},
        pages = {398-400},
          doi = {10.1038/nature08022},
       adsurl = {https://ui.adsabs.harvard.edu/abs/2009Natur.459..398D}
}

@ARTICLE{2012ApJ...756...19D,
       author = {{Deheuvels}, S. and {Garc{\'\i}a}, R.~A. and {Chaplin}, W.~J. and {Basu}, S. and {Antia}, H.~M. and {Appourchaux}, T. and {Benomar}, O. and {Davies}, G.~R. and {Elsworth}, Y. and {Gizon}, L. and {Goupil}, M.~J. and {Reese}, D.~R. and {Regulo}, C. and {Schou}, J. and {Stahn}, T. and {Casagrande}, L. and {Christensen-Dalsgaard}, J. and {Fischer}, D. and {Hekker}, S. and {Kjeldsen}, H. and {Mathur}, S. and {Mosser}, B. and {Pinsonneault}, M. and {Valenti}, J. and {Christiansen}, J.~L. and {Kinemuchi}, K. and {Mullally}, F.},
        title = "{Seismic Evidence for a Rapidly Rotating Core in a Lower-giant-branch Star Observed with Kepler}",
      journal = {\apj},
         year = 2012,
        month = sep,
       volume = {756},
       number = {1},
          eid = {19},
        pages = {19},
          doi = {10.1088/0004-637X/756/1/19},
archivePrefix = {arXiv},
       eprint = {1206.3312},
 primaryClass = {astro-ph.SR},
       adsurl = {https://ui.adsabs.harvard.edu/abs/2012ApJ...756...19D}
}

@ARTICLE{2014A&A...564A..27D,
       author = {{Deheuvels}, S. and {Do{\u{g}}an}, G. and {Goupil}, M.~J. and {Appourchaux}, T. and {Benomar}, O. and {Bruntt}, H. and {Campante}, T.~L. and {Casagrande}, L. and {Ceillier}, T. and {Davies}, G.~R. and {De Cat}, P. and {Fu}, J.~N. and {Garc{\'\i}a}, R.~A. and {Lobel}, A. and {Mosser}, B. and {Reese}, D.~R. and {Regulo}, C. and {Schou}, J. and {Stahn}, T. and {Thygesen}, A.~O. and {Yang}, X.~H. and {Chaplin}, W.~J. and {Christensen-Dalsgaard}, J. and {Eggenberger}, P. and {Gizon}, L. and {Mathis}, S. and {Molenda-{\.Z}akowicz}, J. and {Pinsonneault}, M.},
        title = "{Seismic constraints on the radial dependence of the internal rotation profiles of six Kepler subgiants and young red giants}",
      journal = {\aap},
         year = 2014,
        month = apr,
       volume = {564},
          eid = {A27},
        pages = {A27},
          doi = {10.1051/0004-6361/201322779},
archivePrefix = {arXiv},
       eprint = {1401.3096},
 primaryClass = {astro-ph.SR},
       adsurl = {https://ui.adsabs.harvard.edu/abs/2014A&A...564A..27D}
}

@ARTICLE{2015A&A...580A..96D,
       author = {{Deheuvels}, S. and {Ballot}, J. and {Beck}, P.~G. and {Mosser}, B. and {{\O}stensen}, R. and {Garc{\'\i}a}, R.~A. and {Goupil}, M.~J.},
        title = "{Seismic evidence for a weak radial differential rotation in intermediate-mass core helium burning stars}",
      journal = {\aap},
         year = 2015,
        month = aug,
       volume = {580},
          eid = {A96},
        pages = {A96},
          doi = {10.1051/0004-6361/201526449},
archivePrefix = {arXiv},
       eprint = {1506.02704},
 primaryClass = {astro-ph.SR},
       adsurl = {https://ui.adsabs.harvard.edu/abs/2015A&A...580A..96D}
}

@ARTICLE{2017A&A...605A..75D,
       author = {{Deheuvels}, S. and {Ouazzani}, R.~M. and {Basu}, S.},
        title = "{Near-degeneracy effects on the frequencies of rotationally-split mixed modes in red giants}",
      journal = {\aap},
         year = 2017,
        month = sep,
       volume = {605},
          eid = {A75},
        pages = {A75},
          doi = {10.1051/0004-6361/201730786},
archivePrefix = {arXiv},
       eprint = {1705.10326},
 primaryClass = {astro-ph.SR},
       adsurl = {https://ui.adsabs.harvard.edu/abs/2017A&A...605A..75D}
}

@ARTICLE{2008ApJS..178...89D,
       author = {{Dotter}, Aaron and {Chaboyer}, Brian and {Jevremovi{\'c}}, Darko and {Kostov}, Veselin and {Baron}, E. and {Ferguson}, Jason W.},
        title = "{The Dartmouth Stellar Evolution Database}",
      journal = {\apjs},
         year = 2008,
        month = sep,
       volume = {178},
       number = {1},
        pages = {89-101},
          doi = {10.1086/589654},
archivePrefix = {arXiv},
       eprint = {0804.4473},
 primaryClass = {astro-ph},
       adsurl = {https://ui.adsabs.harvard.edu/abs/2008ApJS..178...89D}
}

@ARTICLE{2009A&A...506...57D,
       author = {{Dupret}, M.-A. and {Belkacem}, K. and {Samadi}, R. and {Montalban}, J. and {Moreira}, O. and {Miglio}, A. and {Godart}, M. and {Ventura}, P. and {Ludwig}, H.-G. and {Grigahc{\`e}ne}, A. and et al.},
        title = "{Theoretical amplitudes and lifetimes of non-radial solar-like oscillations in red giants}",
      journal = {\aap},
         year = 2009,
        month = oct,
       volume = {506},
       number = {1},
        pages = {57-67},
          doi = {10.1051/0004-6361/200911713},
archivePrefix = {arXiv},
       eprint = {0906.3951},
 primaryClass = {astro-ph.SR},
       adsurl = {https://ui.adsabs.harvard.edu/abs/2009A&A...506...57D}
}

@ARTICLE{2005ApJ...623..585F,
       author = {{Ferguson}, Jason W. and {Alexander}, David R. and {Allard}, France and {Barman}, Travis and {Bodnarik}, Julia G. and {Hauschildt}, Peter H. and {Heffner-Wong}, Amanda and {Tamanai}, Akemi},
        title = "{Low-Temperature Opacities}",
      journal = {\apj},
         year = 2005,
        month = apr,
       volume = {623},
       number = {1},
        pages = {585-596},
          doi = {10.1086/428642},
archivePrefix = {arXiv},
       eprint = {astro-ph/0502045},
 primaryClass = {astro-ph},
       adsurl = {https://ui.adsabs.harvard.edu/abs/2005ApJ...623..585F}
}

@ARTICLE{2011ApJ...730...63G,
       author = {{Gai}, Ning and {Basu}, Sarbani and {Chaplin}, William J. and {Elsworth}, Yvonne},
        title = "{An In-depth Study of Grid-based Asteroseismic Analysis}",
      journal = {\apj},
         year = 2011,
        month = apr,
       volume = {730},
       number = {2},
          eid = {63},
        pages = {63},
          doi = {10.1088/0004-637X/730/2/63},
archivePrefix = {arXiv},
       eprint = {1009.3018},
 primaryClass = {astro-ph.SR},
       adsurl = {https://ui.adsabs.harvard.edu/abs/2011ApJ...730...63G}
}

@ARTICLE{2016AJ....151..144G,
       author = {{Garc{\'\i}a P{\'e}rez}, Ana E. and {Allende Prieto}, Carlos and {Holtzman}, Jon A. and {Shetrone}, Matthew and {M{\'e}sz{\'a}ros}, Szabolcs and {Bizyaev}, Dmitry and {Carrera}, Ricardo and {Cunha}, Katia and {Garc{\'\i}a-Hern{\'a}ndez}, D.~A. and {Johnson}, Jennifer A. and {Majewski}, Steven R. and {Nidever}, David L. and {Schiavon}, Ricardo P. and {Shane}, Neville and {Smith}, Verne V. and {Sobeck}, Jennifer and {Troup}, Nicholas and {Zamora}, Olga and {Weinberg}, David H. and {Bovy}, Jo and {Eisenstein}, Daniel J. and {Feuillet}, Diane and {Frinchaboy}, Peter M. and {Hayden}, Michael R. and {Hearty}, Fred R. and {Nguyen}, Duy C. and {O'Connell}, Robert W. and {Pinsonneault}, Marc H. and {Wilson}, John C. and {Zasowski}, Gail},
        title = "{ASPCAP: The APOGEE Stellar Parameter and Chemical Abundances Pipeline}",
      journal = {\aj},
         year = 2016,
        month = jun,
       volume = {151},
       number = {6},
          eid = {144},
        pages = {144},
          doi = {10.3847/0004-6256/151/6/144},
archivePrefix = {arXiv},
       eprint = {1510.07635},
 primaryClass = {astro-ph.SR},
       adsurl = {https://ui.adsabs.harvard.edu/abs/2016AJ....151..144G}
}

@ARTICLE{2013A&A...549A..75G,
       author = {{Goupil}, M.~J. and {Mosser}, B. and {Marques}, J.~P. and {Ouazzani}, R.~M. and {Belkacem}, K. and {Lebreton}, Y. and {Samadi}, R.},
        title = "{Seismic diagnostics for transport of angular momentum in stars. II. Interpreting observed rotational splittings of slowly rotating red giant stars}",
      journal = {\aap},
         year = 2013,
        month = jan,
       volume = {549},
          eid = {A75},
        pages = {A75},
          doi = {10.1051/0004-6361/201220266},
archivePrefix = {arXiv},
       eprint = {1211.1546},
 primaryClass = {astro-ph.SR},
       adsurl = {https://ui.adsabs.harvard.edu/abs/2013A&A...549A..75G}
}

@ARTICLE{1998SSRv...85..161G,
       author = {{Grevesse}, N. and {Sauval}, A.~J.},
        title = "{Standard Solar Composition}",
      journal = {\ssr},
         year = 1998,
        month = may,
       volume = {85},
        pages = {161-174},
          doi = {10.1023/A:1005161325181},
       adsurl = {https://ui.adsabs.harvard.edu/abs/1998SSRv...85..161G}
}

@ARTICLE{2017A&ARv..25....1H,
       author = {{Hekker}, S. and {Christensen-Dalsgaard}, J.},
        title = "{Giant star seismology}",
      journal = {\aapr},
         year = 2017,
        month = jun,
       volume = {25},
       number = {1},
          eid = {1},
        pages = {1},
          doi = {10.1007/s00159-017-0101-x},
archivePrefix = {arXiv},
       eprint = {1609.07487},
 primaryClass = {astro-ph.SR},
       adsurl = {https://ui.adsabs.harvard.edu/abs/2017A&ARv..25....1H}
}

@ARTICLE{2000A&A...360..952H,
       author = {{Herwig}, F.},
        title = "{The evolution of AGB stars with convective overshoot}",
      journal = {\aap},
         year = 2000,
        month = aug,
       volume = {360},
        pages = {952-968},
          doi = {10.48550/arXiv.astro-ph/0007139},
archivePrefix = {arXiv},
       eprint = {astro-ph/0007139},
 primaryClass = {astro-ph},
       adsurl = {https://ui.adsabs.harvard.edu/abs/2000A&A...360..952H}
}

@ARTICLE{2011ApJ...743..143H,
       author = {{Huber}, D. and {Bedding}, T.~R. and {Stello}, D. and {Hekker}, S. and {Mathur}, S. and {Mosser}, B. and {Verner}, G.~A. and {Bonanno}, A. and {Buzasi}, D.~L. and {Campante}, T.~L. and {Elsworth}, Y.~P. and {Hale}, S.~J. and {Kallinger}, T. and {Silva Aguirre}, V. and {Chaplin}, W.~J. and {De Ridder}, J. and {Garc{\'\i}a}, R.~A. and {Appourchaux}, T. and {Frandsen}, S. and {Houdek}, G. and {Molenda-{\.Z}akowicz}, J. and {Monteiro}, M.~J.~P.~F.~G. and {Christensen-Dalsgaard}, J. and {Gilliland}, R.~L. and {Kawaler}, S.~D. and {Kjeldsen}, H. and {Broomhall}, A.~M. and {Corsaro}, E. and {Salabert}, D. and {Sanderfer}, D.~T. and {Seader}, S.~E. and {Smith}, J.~C.},
        title = "{Testing Scaling Relations for Solar-like Oscillations from the Main Sequence to Red Giants Using Kepler Data}",
      journal = {\apj},
         year = 2011,
        month = dec,
       volume = {743},
       number = {2},
          eid = {143},
        pages = {143},
          doi = {10.1088/0004-637X/743/2/143},
archivePrefix = {arXiv},
       eprint = {1109.3460},
 primaryClass = {astro-ph.SR},
       adsurl = {https://ui.adsabs.harvard.edu/abs/2011ApJ...743..143H}
}

@ARTICLE{2017ApJ...844..102H,
       author = {{Huber}, Daniel and {Zinn}, Joel and {Bojsen-Hansen}, Mathias and {Pinsonneault}, Marc and {Sahlholdt}, Christian and {Serenelli}, Aldo and {Silva Aguirre}, Victor and {Stassun}, Keivan and {Stello}, Dennis and {Tayar}, Jamie and {Bastien}, Fabienne and {Bedding}, Timothy R. and {Buchhave}, Lars A. and {Chaplin}, William J. and {Davies}, Guy R. and {Garc{\'\i}a}, Rafael A. and {Latham}, David W. and {Mathur}, Savita and {Mosser}, Benoit and {Sharma}, Sanjib},
        title = "{Asteroseismology and Gaia: Testing Scaling Relations Using 2200 Kepler  Stars with TGAS Parallaxes}",
      journal = {\apj},
         year = 2017,
        month = aug,
       volume = {844},
       number = {2},
          eid = {102},
        pages = {102},
          doi = {10.3847/1538-4357/aa75ca},
archivePrefix = {arXiv},
       eprint = {1705.04697},
 primaryClass = {astro-ph.SR},
       adsurl = {https://ui.adsabs.harvard.edu/abs/2017ApJ...844..102H}
}

@ARTICLE{2023arXiv230703237H,
       author = {{Huber}, Daniel and {Pinsonneault}, Marc and {Beck}, Paul and {Bedding}, Timothy R. and {Bland-Hawthorn}, Joss and {Breton}, Sylvain N. and {Bugnet}, Lisa and {Chaplin}, William J. and {Garcia}, Rafael A. and {Grunblatt}, Samuel K. and {Guzik}, Joyce A. and {Hekker}, Saskia and {Kawaler}, Steven D. and {Mathis}, Stephane and {Mathur}, Savita and {Metcalfe}, Travis and {Mosser}, Benoit and {Ness}, Melissa K. and {Piro}, Anthony L. and {Serenelli}, Aldo and {Sharma}, Sanjib and {Soderblom}, David R. and {Stassun}, Keivan G. and {Stello}, Dennis and {Tayar}, Jamie and {van Belle}, Gerard T. and {Zinn}, Joel C.},
        title = "{Asteroseismology with the Roman Galactic Bulge Time-Domain Survey}",
      journal = {arXiv e-prints},
         year = 2023,
        month = jul,
          eid = {arXiv:2307.03237},
        pages = {arXiv:2307.03237},
          doi = {10.48550/arXiv.2307.03237},
archivePrefix = {arXiv},
       eprint = {2307.03237},
 primaryClass = {astro-ph.IM},
       adsurl = {https://ui.adsabs.harvard.edu/abs/2023arXiv230703237H}
}

@ARTICLE{1996ApJ...464..943I,
       author = {{Iglesias}, Carlos A. and {Rogers}, Forrest J.},
        title = "{Updated Opal Opacities}",
      journal = {\apj},
         year = 1996,
        month = jun,
       volume = {464},
        pages = {943},
          doi = {10.1086/177381},
       adsurl = {https://ui.adsabs.harvard.edu/abs/1996ApJ...464..943I}
}

@BOOK{2013sse..book.....K,
       author = {{Kippenhahn}, Rudolf and {Weigert}, Alfred and {Weiss}, Achim},
        title = "{Stellar Structure and Evolution}",
         year = 2013,
          doi = {10.1007/978-3-642-30304-3},
       adsurl = {https://ui.adsabs.harvard.edu/abs/2013sse..book.....K}
}

@ARTICLE{2023ApJ...954..152K,
       author = {{Kuszlewicz}, James S. and {Hon}, Marc and {Huber}, Daniel},
        title = "{Mixed-mode Ensemble Asteroseismology of Low-luminosity Kepler Red Giants}",
      journal = {\apj},
         year = 2023,
        month = sep,
       volume = {954},
       number = {2},
          eid = {152},
        pages = {152},
          doi = {10.3847/1538-4357/ace598},
archivePrefix = {arXiv},
       eprint = {2307.06482},
 primaryClass = {astro-ph.SR},
       adsurl = {https://ui.adsabs.harvard.edu/abs/2023ApJ...954..152K}
}

@ARTICLE{2018MNRAS.475..981L,
       author = {{Li}, Tanda and {Bedding}, Timothy R. and {Huber}, Daniel and {Ball}, Warrick H. and {Stello}, Dennis and {Murphy}, Simon J. and {Bland-Hawthorn}, Joss},
        title = "{Modelling Kepler red giants in eclipsing binaries: calibrating the mixing-length parameter with asteroseismology}",
      journal = {\mnras},
         year = 2018,
        month = mar,
       volume = {475},
       number = {1},
        pages = {981-998},
          doi = {10.1093/mnras/stx3079},
archivePrefix = {arXiv},
       eprint = {1712.01424},
 primaryClass = {astro-ph.SR},
       adsurl = {https://ui.adsabs.harvard.edu/abs/2018MNRAS.475..981L}
}

@ARTICLE{2022ApJ...927..167L,
       author = {{Li}, Tanda and {Li}, Yaguang and {Bi}, Shaolan and {Bedding}, Timothy R. and {Davies}, Guy and {Du}, Minghao},
        title = "{Asteroseismology of 3642 Kepler Red Giants: Correcting the Scaling Relations Based on Detailed Modeling}",
      journal = {\apj},
         year = 2022,
        month = mar,
       volume = {927},
       number = {2},
          eid = {167},
        pages = {167},
          doi = {10.3847/1538-4357/ac4fbf},
archivePrefix = {arXiv},
       eprint = {2201.09577},
 primaryClass = {astro-ph.SR},
       adsurl = {https://ui.adsabs.harvard.edu/abs/2022ApJ...927..167L}
}

@ARTICLE{LiG2024,
       author = {{Li}, Gang and {Deheuvels}, S{\'e}bastien and {Ballot}, J{\'e}r{\^o}me},
        title = "{Asteroseismic measurement of core and envelope rotation rates for 2006 red giant branch stars}",
      journal = {\aap},
         year = 2024,
        month = aug,
       volume = {688},
          eid = {A184},
        pages = {A184},
          doi = {10.1051/0004-6361/202449882},
archivePrefix = {arXiv},
       eprint = {2405.12116},
 primaryClass = {astro-ph.SR},
       adsurl = {https://ui.adsabs.harvard.edu/abs/2024A&A...688A.184L}
}

@ARTICLE{2024ApJ...974...77L,
       author = {{Li}, Yaguang and {Bedding}, Timothy R. and {Huber}, Daniel and {Stello}, Dennis and {van Saders}, Jennifer and {Zhou}, Yixiao and {Crawford}, Courtney L. and {Joyce}, Meridith and {Li}, Tanda and {Murphy}, Simon J. and {Sreenivas}, K.~R.},
        title = "{Realistic Uncertainties for Fundamental Properties of Asteroseismic Red Giants and the Interplay between Mixing Length, Metallicity, and <inline-formula> <mml:math overflow=``scroll''><mml:msub><mml:mrow><mml:mi>{\ensuremath{\nu}}</mml:mi></mml:mrow><mml:mrow><mml:mi>max</mml:mi></mml:mrow></mml:msub></mml:math> </inline-formula>}",
      journal = {\apj},
         year = 2024,
        month = oct,
       volume = {974},
       number = {1},
          eid = {77},
        pages = {77},
          doi = {10.3847/1538-4357/ad6c3e},
archivePrefix = {arXiv},
       eprint = {2407.09967},
 primaryClass = {astro-ph.SR},
       adsurl = {https://ui.adsabs.harvard.edu/abs/2024ApJ...974...77L}
}

@ARTICLE{2010ApJ...718.1378M,
       author = {{Magic}, Z. and {Serenelli}, A. and {Weiss}, A. and {Chaboyer}, B.},
        title = "{On Using the Color-Magnitude Diagram Morphology of M67 to Test Solar Abundances}",
      journal = {\apj},
         year = 2010,
        month = aug,
       volume = {718},
       number = {2},
        pages = {1378-1387},
          doi = {10.1088/0004-637X/718/2/1378},
archivePrefix = {arXiv},
       eprint = {1004.3308},
 primaryClass = {astro-ph.SR},
       adsurl = {https://ui.adsabs.harvard.edu/abs/2010ApJ...718.1378M}
}

@ARTICLE{2015A&A...573A..89M,
       author = {{Magic}, Z. and {Weiss}, A. and {Asplund}, M.},
        title = "{The Stagger-grid: A grid of 3D stellar atmosphere models. III. The relation to mixing length convection theory}",
      journal = {\aap},
         year = 2015,
        month = jan,
       volume = {573},
          eid = {A89},
        pages = {A89},
          doi = {10.1051/0004-6361/201423760},
archivePrefix = {arXiv},
       eprint = {1403.1062},
 primaryClass = {astro-ph.SR},
       adsurl = {https://ui.adsabs.harvard.edu/abs/2015A&A...573A..89M}
}

@ARTICLE{2014ApJS..214...27M,
       author = {{Metcalfe}, T.~S. and {Creevey}, O.~L. and {Do{\u{g}}an}, G. and {Mathur}, S. and {Xu}, H. and {Bedding}, T.~R. and {Chaplin}, W.~J. and {Christensen-Dalsgaard}, J. and {Karoff}, C. and {Trampedach}, R. and {Benomar}, O. and {Brown}, B.~P. and {Buzasi}, D.~L. and {Campante}, T.~L. and {{\c{C}}elik}, Z. and {Cunha}, M.~S. and {Davies}, G.~R. and {Deheuvels}, S. and {Derekas}, A. and {Di Mauro}, M.~P. and {Garc{\'\i}a}, R.~A. and {Guzik}, J.~A. and {Howe}, R. and {MacGregor}, K.~B. and {Mazumdar}, A. and {Montalb{\'a}n}, J. and {Monteiro}, M.~J.~P.~F.~G. and {Salabert}, D. and {Serenelli}, A. and {Stello}, D. and {Ste\&{\textcommabelow s}acute} and {licki}, M. and {Suran}, M.~D. and {Y{\i}ld{\i}z}, M. and {Aksoy}, C. and {Elsworth}, Y. and {Gruberbauer}, M. and {Guenther}, D.~B. and {Lebreton}, Y. and {Molaverdikhani}, K. and {Pricopi}, D. and {Simoniello}, R. and {White}, T.~R.},
        title = "{Properties of 42 Solar-type Kepler Targets from the Asteroseismic Modeling Portal}",
      journal = {\apjs},
         year = 2014,
        month = oct,
       volume = {214},
       number = {2},
          eid = {27},
        pages = {27},
          doi = {10.1088/0067-0049/214/2/27},
archivePrefix = {arXiv},
       eprint = {1402.3614},
 primaryClass = {astro-ph.SR},
       adsurl = {https://ui.adsabs.harvard.edu/abs/2014ApJS..214...27M}
}

@ARTICLE{2011A&A...532A..86M,
       author = {{Mosser}, B. and {Barban}, C. and {Montalb{\'a}n}, J. and {Beck}, P.~G. and {Miglio}, A. and {Belkacem}, K. and {Goupil}, M.~J. and {Hekker}, S. and {De Ridder}, J. and {Dupret}, M.~A. and {Elsworth}, Y. and {Noels}, A. and {Baudin}, F. and {Michel}, E. and {Samadi}, R. and {Auvergne}, M. and {Baglin}, A. and {Catala}, C.},
        title = "{Mixed modes in red-giant stars observed with CoRoT}",
      journal = {\aap},
         year = 2011,
        month = aug,
       volume = {532},
          eid = {A86},
        pages = {A86},
          doi = {10.1051/0004-6361/201116825},
archivePrefix = {arXiv},
       eprint = {1105.6113},
 primaryClass = {astro-ph.SR},
       adsurl = {https://ui.adsabs.harvard.edu/abs/2011A&A...532A..86M}
}

@ARTICLE{2012A&A...548A..10M,
       author = {{Mosser}, B. and {Goupil}, M.~J. and {Belkacem}, K. and {Marques}, J.~P. and {Beck}, P.~G. and {Bloemen}, S. and {De Ridder}, J. and {Barban}, C. and {Deheuvels}, S. and {Elsworth}, Y. and {Hekker}, S. and {Kallinger}, T. and {Ouazzani}, R.~M. and {Pinsonneault}, M. and {Samadi}, R. and {Stello}, D. and {Garc{\'\i}a}, R.~A. and {Klaus}, T.~C. and {Li}, J. and {Mathur}, S. and {Morris}, R.~L.},
        title = "{Spin down of the core rotation in red giants}",
      journal = {\aap},
         year = 2012,
        month = dec,
       volume = {548},
          eid = {A10},
        pages = {A10},
          doi = {10.1051/0004-6361/201220106},
archivePrefix = {arXiv},
       eprint = {1209.3336},
 primaryClass = {astro-ph.SR},
       adsurl = {https://ui.adsabs.harvard.edu/abs/2012A&A...548A..10M}
}

@ARTICLE{Mosser2012,
       author = {{Mosser}, B. and {Goupil}, M.~J. and {Belkacem}, K. and {Michel}, E. and {Stello}, D. and {Marques}, J.~P. and {Elsworth}, Y. and {Barban}, C. and {Beck}, P.~G. and {Bedding}, T.~R. and et al.},
        title = "{Probing the core structure and evolution of red giants using gravity-dominated mixed modes observed with Kepler}",
      journal = {\aap},
         year = 2012,
        month = apr,
       volume = {540},
          eid = {A143},
        pages = {A143},
          doi = {10.1051/0004-6361/201118519},
archivePrefix = {arXiv},
       eprint = {1203.0689},
 primaryClass = {astro-ph.SR},
       adsurl = {https://ui.adsabs.harvard.edu/abs/2012A&A...540A.143M}
}

@ARTICLE{2014A&A...572L...5M,
       author = {{Mosser}, B. and {Benomar}, O. and {Belkacem}, K. and {Goupil}, M.~J. and {Lagarde}, N. and {Michel}, E. and {Lebreton}, Y. and {Stello}, D. and {Vrard}, M. and {Barban}, C. and {Bedding}, T.~R. and {Deheuvels}, S. and {Chaplin}, W.~J. and {De Ridder}, J. and {Elsworth}, Y. and {Montalban}, J. and {Noels}, A. and {Ouazzani}, R.~M. and {Samadi}, R. and {White}, T.~R. and {Kjeldsen}, H.},
        title = "{Mixed modes in red giants: a window on stellar evolution}",
      journal = {\aap},
         year = 2014,
        month = dec,
       volume = {572},
          eid = {L5},
        pages = {L5},
          doi = {10.1051/0004-6361/201425039},
archivePrefix = {arXiv},
       eprint = {1411.1082},
 primaryClass = {astro-ph.SR},
       adsurl = {https://ui.adsabs.harvard.edu/abs/2014A&A...572L...5M}
}

@ARTICLE{2015A&A...584A..50M,
       author = {{Mosser}, B. and {Vrard}, M. and {Belkacem}, K. and {Deheuvels}, S. and {Goupil}, M.~J.},
        title = "{Period spacings in red giants. I. Disentangling rotation and revealing core structure discontinuities}",
      journal = {\aap},
         year = 2015,
        month = dec,
       volume = {584},
          eid = {A50},
        pages = {A50},
          doi = {10.1051/0004-6361/201527075},
archivePrefix = {arXiv},
       eprint = {1509.06193},
 primaryClass = {astro-ph.SR},
       adsurl = {https://ui.adsabs.harvard.edu/abs/2015A&A...584A..50M}
}

@ARTICLE{2017A&A...600A...1M,
       author = {{Mosser}, B. and {Pin{\c{c}}on}, C. and {Belkacem}, K. and {Takata}, M. and {Vrard}, M.},
        title = "{Period spacings in red giants. III. Coupling factors of mixed modes}",
      journal = {\aap},
         year = 2017,
        month = apr,
       volume = {600},
          eid = {A1},
        pages = {A1},
          doi = {10.1051/0004-6361/201630053},
archivePrefix = {arXiv},
       eprint = {1612.08453},
 primaryClass = {astro-ph.SR},
       adsurl = {https://ui.adsabs.harvard.edu/abs/2017A&A...600A...1M}
}

@ARTICLE{2021MNRAS.505.2336M,
       author = {{Murphy}, Simon J. and {Li}, Tanda and {Sekaran}, Sanjay and {Bedding}, Timothy R. and {Yu}, Jie and {Tkachenko}, Andrew and {Colman}, Isabel and {Huber}, Daniel and {Hey}, Daniel and {Baratashvili}, Tinatin and {Janssens}, Soetkin},
        title = "{A binary with a {\ensuremath{\delta}} Scuti star and an oscillating red giant: orbit and asteroseismology of KIC 9773821}",
      journal = {\mnras},
         year = 2021,
        month = aug,
       volume = {505},
       number = {2},
        pages = {2336-2348},
          doi = {10.1093/mnras/stab1436},
archivePrefix = {arXiv},
       eprint = {2105.13577},
 primaryClass = {astro-ph.SR},
       adsurl = {https://ui.adsabs.harvard.edu/abs/2021MNRAS.505.2336M}
}

@ARTICLE{Ong2024,
       author = {{Ong}, J.~M. Joel},
        title = "{Red Giant Rotational Inversion Kernels Need Nonlinear Surface Corrections}",
      journal = {\apj},
         year = 2024,
        month = jan,
       volume = {960},
       number = {1},
          eid = {2},
        pages = {2},
          doi = {10.3847/1538-4357/ad0cac},
archivePrefix = {arXiv},
       eprint = {2311.06990},
 primaryClass = {astro-ph.SR},
       adsurl = {https://ui.adsabs.harvard.edu/abs/2024ApJ...960....2O}
}

@ARTICLE{2011ApJS..192....3P,
       author = {{Paxton}, Bill and {Bildsten}, Lars and {Dotter}, Aaron and {Herwig}, Falk and {Lesaffre}, Pierre and {Timmes}, Frank},
        title = "{Modules for Experiments in Stellar Astrophysics (MESA)}",
      journal = {\apjs},
         year = 2011,
        month = jan,
       volume = {192},
       number = {1},
          eid = {3},
        pages = {3},
          doi = {10.1088/0067-0049/192/1/3},
archivePrefix = {arXiv},
       eprint = {1009.1622},
 primaryClass = {astro-ph.SR},
       adsurl = {https://ui.adsabs.harvard.edu/abs/2011ApJS..192....3P}
}

@ARTICLE{2013ApJS..208....4P,
       author = {{Paxton}, Bill and {Cantiello}, Matteo and {Arras}, Phil and {Bildsten}, Lars and {Brown}, Edward F. and {Dotter}, Aaron and {Mankovich}, Christopher and {Montgomery}, M.~H. and {Stello}, Dennis and {Timmes}, F.~X. and {Townsend}, Richard},
        title = "{Modules for Experiments in Stellar Astrophysics (MESA): Planets, Oscillations, Rotation, and Massive Stars}",
      journal = {\apjs},
         year = 2013,
        month = sep,
       volume = {208},
       number = {1},
          eid = {4},
        pages = {4},
          doi = {10.1088/0067-0049/208/1/4},
archivePrefix = {arXiv},
       eprint = {1301.0319},
 primaryClass = {astro-ph.SR},
       adsurl = {https://ui.adsabs.harvard.edu/abs/2013ApJS..208....4P}
}

@ARTICLE{2015ApJS..220...15P,
       author = {{Paxton}, Bill and {Marchant}, Pablo and {Schwab}, Josiah and {Bauer}, Evan B. and {Bildsten}, Lars and {Cantiello}, Matteo and {Dessart}, Luc and {Farmer}, R. and {Hu}, H. and {Langer}, N. and {Townsend}, R.~H.~D. and {Townsley}, Dean M. and {Timmes}, F.~X.},
        title = "{Modules for Experiments in Stellar Astrophysics (MESA): Binaries, Pulsations, and Explosions}",
      journal = {\apjs},
         year = 2015,
        month = sep,
       volume = {220},
       number = {1},
          eid = {15},
        pages = {15},
          doi = {10.1088/0067-0049/220/1/15},
archivePrefix = {arXiv},
       eprint = {1506.03146},
 primaryClass = {astro-ph.SR},
       adsurl = {https://ui.adsabs.harvard.edu/abs/2015ApJS..220...15P}
}

@ARTICLE{2018ApJS..234...34P,
       author = {{Paxton}, Bill and {Schwab}, Josiah and {Bauer}, Evan B. and {Bildsten}, Lars and {Blinnikov}, Sergei and {Duffell}, Paul and {Farmer}, R. and {Goldberg}, Jared A. and {Marchant}, Pablo and {Sorokina}, Elena and {Thoul}, Anne and {Townsend}, Richard H.~D. and {Timmes}, F.~X.},
        title = "{Modules for Experiments in Stellar Astrophysics (MESA): Convective Boundaries, Element Diffusion, and Massive Star Explosions}",
      journal = {\apjs},
         year = 2018,
        month = feb,
       volume = {234},
       number = {2},
          eid = {34},
        pages = {34},
          doi = {10.3847/1538-4365/aaa5a8},
archivePrefix = {arXiv},
       eprint = {1710.08424},
 primaryClass = {astro-ph.SR},
       adsurl = {https://ui.adsabs.harvard.edu/abs/2018ApJS..234...34P}
}

@ARTICLE{perez2016,
       author = {{P{\'e}rez Hern{\'a}ndez}, F. and {Garc{\'\i}a}, R.~A. and {Corsaro}, E. and {Triana}, S.~A. and {De Ridder}, J.},
        title = "{Asteroseismology of 19 low-luminosity red giant stars from Kepler}",
      journal = {\aap},
         year = 2016,
        month = jun,
       volume = {591},
          eid = {A99},
        pages = {A99},
          doi = {10.1051/0004-6361/201628311},
archivePrefix = {arXiv},
       eprint = {1605.01567},
 primaryClass = {astro-ph.SR},
       adsurl = {https://ui.adsabs.harvard.edu/abs/2016A&A...591A..99P}
}

@ARTICLE{2025ExA....59...26R,
       author = {{Rauer}, Heike and {Aerts}, Conny and {Cabrera}, Juan and {Deleuil}, Magali and {Erikson}, Anders and {Gizon}, Laurent and {Goupil}, Mariejo and {Heras}, Ana and {Walloschek}, Thomas and {Lorenzo-Alvarez}, Jose and {Marliani}, Filippo and {Martin-Garcia}, C{\'e}sar and {Mas-Hesse}, J. Miguel and {O'Rourke}, Laurence and {Osborn}, Hugh and {Pagano}, Isabella and {Piotto}, Giampaolo and {Pollacco}, Don and {Ragazzoni}, Roberto and {Ramsay}, Gavin and {Udry}, St{\'e}phane and {Appourchaux}, Thierry and {Benz}, Willy and {Brandeker}, Alexis and {G{\"u}del}, Manuel and {Janot-Pacheco}, Eduardo and {Kabath}, Petr and {Kjeldsen}, Hans and {Min}, Michiel and {Santos}, Nuno and {Smith}, Alan and {Suarez}, Juan-Carlos and {Werner}, Stephanie C. and {Aboudan}, Alessio and {Abreu}, Manuel and {Acu{\~n}a}, Lorena and {Adams}, Moritz and {Adibekyan}, Vardan and {Affer}, Laura and {Agneray}, Fran{\c{c}}ois and {Agnor}, Craig and {Aguirre B{\o}rsen-Koch}, Victor and {Ahmed}, Saad and {Aigrain}, Suzanne and {Al-Bahlawan}, Ashraf and {Alcacera Gil}, Ma de los Angeles and {Alei}, Eleonora and {Alencar}, Silvia and {Alexander}, Richard and {Alfonso-Garz{\'o}n}, Julia and {Alibert}, Yann and {Allende Prieto}, Carlos and {Almeida}, Leonardo and {Alonso Sobrino}, Roi and {Altavilla}, Giuseppe and {Althaus}, Christian and {Alvarez Trujillo}, Luis Alonso and {Amarsi}, Anish and {Ammler-von Eiff}, Matthias and {Am{\^o}res}, Eduardo and {Andrade}, Laerte and {Antoniadis-Karnavas}, Alexandros and {Ant{\'o}nio}, Carlos and {Aparicio del Moral}, Beatriz and {Appolloni}, Matteo and {Arena}, Claudio and {Armstrong}, David and {Aroca Aliaga}, Jose and {Asplund}, Martin and {Audenaert}, Jeroen and {Auricchio}, Natalia and {Avelino}, Pedro and {Baeke}, Ann and {Bailli{\'e}}, Kevin and {Balado}, Ana and {Ballber Balaguer{\'o}}, Pau and {Balestra}, Andrea and {Ball}, Warrick and {Ballans}, Herve and {Ballot}, Jerome and {Barban}, Caroline and {Barbary}, Ga{\"e}le and {Barbieri}, Mauro and {Barcel{\'o} Forteza}, Sebasti{\`a} and {Barker}, Adrian and {Barklem}, Paul and {Barnes}, Sydney and {Barrado Navascues}, David and {Barragan}, Oscar and {Baruteau}, Cl{\'e}ment and {Basu}, Sarbani and {Baudin}, Frederic and {Baumeister}, Philipp and {Bayliss}, Daniel and {Bazot}, Michael and {Beck}, Paul G. and {Belkacem}, Kevin and {Bellinger}, Earl and {Benatti}, Serena and {Benomar}, Othman and {B{\'e}rard}, Diane and {Bergemann}, Maria and {Bergomi}, Maria and {Bernardo}, Pierre and {Biazzo}, Katia and {Bignamini}, Andrea and {Bigot}, Lionel and {Billot}, Nicolas and {Binet}, Martin and {Biondi}, David and {Biondi}, Federico and {Birch}, Aaron C. and {Bitsch}, Bertram and {Bluhm Ceballos}, Paz Victoria and {B{\'o}di}, Attila and {Bogn{\'a}r}, Zs{\'o}fia and {Boisse}, Isabelle and {Bolmont}, Emeline and {Bonanno}, Alfio and {Bonavita}, Mariangela and {Bonfanti}, Andrea and {Bonfils}, Xavier and {Bonito}, Rosaria and {Bonomo}, Aldo Stefano and {B{\"o}rner}, Anko and {Boro Saikia}, Sudeshna and {Borreguero Mart{\'\i}n}, Elisa and {Borsa}, Francesco and {Borsato}, Luca and {Bossini}, Diego and {Bouchy}, Francois and {Bou{\'e}}, Gwena{\"e}l and {Boufleur}, Rodrigo and {Boumier}, Patrick and {Bourrier}, Vincent and {Bowman}, Dominic M. and {Bozzo}, Enrico and {Bradley}, Louisa and {Bray}, John and {Bressan}, Alessandro and {Breton}, Sylvain and {Brienza}, Daniele and {Brito}, Ana and {Brogi}, Matteo and {Brown}, Beverly and {Brown}, David J.~A. and {Brun}, Allan Sacha and {Bruno}, Giovanni and {Bruns}, Michael and {Buchhave}, Lars A. and {Bugnet}, Lisa and {Buldgen}, Ga{\"e}l and {Burgess}, Patrick and {Busatta}, Andrea and {Busso}, Giorgia and {Buzasi}, Derek and {Caballero}, Jos{\'e} A. and {Cabral}, Alexandre and {Cabrero Gomez}, Juan-Francisco and {Calderone}, Flavia and {Cameron}, Robert and {Cameron}, Andrew and {Campante}, Tiago and {Campos Gestal}, N{\'e}stor and {Canto Martins}, Bruno Leonardo and {Cara}, Christophe and {Carone}, Ludmila and {Carrasco}, Josep Manel and {Casagrande}, Luca and {Casewell}, Sarah L. and {Cassisi}, Santi and {Castellani}, Marco and {Castro}, Matthieu and {Catala}, Claude and {Catal{\'a}n Fern{\'a}ndez}, Irene and {Catelan}, M{\'a}rcio and {Cegla}, Heather and {Cerruti}, Chiara and {Cessa}, Virginie and {Chadid}, Merieme and {Chaplin}, William and {Charpinet}, Stephane and {Chiappini}, Cristina and {Chiarucci}, Simone and {Chiavassa}, Andrea and {Chinellato}, Simonetta and {Chirulli}, Giovanni and {Christensen-Dalsgaard}, J{\o}rgen and {Church}, Ross and {Claret}, Antonio and {Clarke}, Cathie and {Claudi}, Riccardo and {Clermont}, Lionel and {Coelho}, Hugo and {Coelho}, Joao and {Cogato}, Fabrizio and {Colom{\'e}}, Josep and {Condamin}, Mathieu and {Conde Garc{\'\i}a}, Fernando and {Conseil}, Simon},
        title = "{The PLATO mission}",
      journal = {Experimental Astronomy},
         year = 2025,
        month = jun,
       volume = {59},
       number = {3},
          eid = {26},
        pages = {26},
          doi = {10.1007/s10686-025-09985-9},
archivePrefix = {arXiv},
       eprint = {2406.05447},
 primaryClass = {astro-ph.IM},
       adsurl = {https://ui.adsabs.harvard.edu/abs/2025ExA....59...26R}
}

@ARTICLE{2002ApJ...576.1064R,
       author = {{Rogers}, F.~J. and {Nayfonov}, A.},
        title = "{Updated and Expanded OPAL Equation-of-State Tables: Implications for Helioseismology}",
      journal = {\apj},
         year = 2002,
        month = sep,
       volume = {576},
       number = {2},
        pages = {1064-1074},
          doi = {10.1086/341894},
       adsurl = {https://ui.adsabs.harvard.edu/abs/2002ApJ...576.1064R}
}

@BOOK{2005essp.book.....S,
       author = {{Salaris}, Maurizio and {Cassisi}, Santi},
        title = "{Evolution of Stars and Stellar Populations}",
         year = 2005,
          doi = {10.1002/0470033452},
       adsurl = {https://ui.adsabs.harvard.edu/abs/2005essp.book.....S}
}

@ARTICLE{1980ApJS...43..469T,
       author = {{Tassoul}, M.},
        title = "{Asymptotic approximations for stellar nonradial pulsations.}",
      journal = {\apjs},
         year = 1980,
        month = aug,
       volume = {43},
        pages = {469-490},
          doi = {10.1086/190678},
       adsurl = {https://ui.adsabs.harvard.edu/abs/1980ApJS...43..469T}
}

@ARTICLE{2014AJ....148...85T,
       author = {{Thompson}, B. and {Frinchaboy}, P. and {Kinemuchi}, K. and {Sarajedini}, A. and {Cohen}, R.},
        title = "{WIYN Open Cluster Study. LXII. Comparison of Isochrone Systems using Deep Multi-band Photometry of M35}",
      journal = {\aj},
         year = 2014,
        month = nov,
       volume = {148},
       number = {5},
          eid = {85},
        pages = {85},
          doi = {10.1088/0004-6256/148/5/85},
archivePrefix = {arXiv},
       eprint = {1408.1684},
 primaryClass = {astro-ph.SR},
       adsurl = {https://ui.adsabs.harvard.edu/abs/2014AJ....148...85T}
}

@ARTICLE{2014MNRAS.445.4366T,
       author = {{Trampedach}, Regner and {Stein}, Robert F. and {Christensen-Dalsgaard}, J{\o}rgen and {Nordlund}, {\r{A}}ke and {Asplund}, Martin},
        title = "{Improvements to stellar structure models, based on a grid of 3D convection simulations - II. Calibrating the mixing-length formulation}",
      journal = {\mnras},
         year = 2014,
        month = dec,
       volume = {445},
       number = {4},
        pages = {4366-4384},
          doi = {10.1093/mnras/stu2084},
archivePrefix = {arXiv},
       eprint = {1410.1559},
 primaryClass = {astro-ph.SR},
       adsurl = {https://ui.adsabs.harvard.edu/abs/2014MNRAS.445.4366T}
}

@ARTICLE{2018ApJ...858...28V,
       author = {{Viani}, Lucas S. and {Basu}, Sarbani and {Ong J.}, M. Joel and {Bonaca}, Ana and {Chaplin}, William J.},
        title = "{Investigating the Metallicity-Mixing-length Relation}",
      journal = {\apj},
         year = 2018,
        month = may,
       volume = {858},
       number = {1},
          eid = {28},
        pages = {28},
          doi = {10.3847/1538-4357/aab7eb},
archivePrefix = {arXiv},
       eprint = {1803.05924},
 primaryClass = {astro-ph.SR},
       adsurl = {https://ui.adsabs.harvard.edu/abs/2018ApJ...858...28V}
}

@ARTICLE{2023ApJ...953..182W,
       author = {{Wang}, Yingxiang and {Li}, Tanda and {Bi}, Shaolan and {Bedding}, Timothy R. and {Li}, Yaguang},
        title = "{Asteroseismic Modeling of 1153 Kepler Red Giant Branch Stars: Improved Stellar Parameters with Gravity-mode Period Spacings and Luminosity Constraints}",
      journal = {\apj},
         year = 2023,
        month = aug,
       volume = {953},
       number = {2},
          eid = {182},
        pages = {182},
          doi = {10.3847/1538-4357/ace4c9},
archivePrefix = {arXiv},
       eprint = {2307.03397},
 primaryClass = {astro-ph.SR},
       adsurl = {https://ui.adsabs.harvard.edu/abs/2023ApJ...953..182W}
}

@ARTICLE{2025ApJ...987..181W,
       author = {{Weiss}, Trevor J. and {Downing}, Noah J. and {Pinsonneault}, Marc H. and {Zinn}, Joel C. and {Stello}, Dennis and {Bedding}, Timothy R. and {Cao}, Kaili and {Hon}, Marc and {Reyes}, Claudia and {Gaudi}, B. Scott and {Wilson}, Robert F. and {Huber}, Daniel and {Sharma}, Sanjib},
        title = "{Modeling Asteroseismic Yields for the Roman Galactic Bulge Time-domain Survey}",
      journal = {\apj},
         year = 2025,
        month = jul,
       volume = {987},
       number = {2},
          eid = {181},
        pages = {181},
          doi = {10.3847/1538-4357/adde5b},
archivePrefix = {arXiv},
       eprint = {2503.04999},
 primaryClass = {astro-ph.SR},
       adsurl = {https://ui.adsabs.harvard.edu/abs/2025ApJ...987..181W}
}

@ARTICLE{2018ApJ...855...16Z,
       author = {{Zhang}, Xinyi and {Wu}, Tao and {Li}, Yan},
        title = "{Frequency Identification and Asteroseismic Analysis of the Red Giant KIC 9145955: Fundamental Parameters and Helium Core Size}",
      journal = {\apj},
         year = 2018,
        month = mar,
       volume = {855},
       number = {1},
          eid = {16},
        pages = {16},
          doi = {10.3847/1538-4357/aaaabb},
archivePrefix = {arXiv},
       eprint = {1802.01320},
 primaryClass = {astro-ph.SR},
       adsurl = {https://ui.adsabs.harvard.edu/abs/2018ApJ...855...16Z}
}

@ARTICLE{2020MNRAS.494..511Z,
       author = {{Zhang}, Xinyi and {Li}, Yan and {Wu}, Tao and {Su}, Jie},
        title = "{Determining the size of the helium core of KIC 9970396 using asteroseismology: a red giant approaching the red giant bump}",
      journal = {\mnras},
         year = 2020,
        month = may,
       volume = {494},
       number = {1},
        pages = {511-528},
          doi = {10.1093/mnras/staa667},
       adsurl = {https://ui.adsabs.harvard.edu/abs/2020MNRAS.494..511Z}
}

@ARTICLE{2022ApJ...931...64Z,
       author = {{Zhang}, Xinyi and {Cai}, Tao and {Li}, Yan and {Wu}, Tao},
        title = "{Determining the Age for the Red Giants KIC 9145955 and KIC 9970396 by Gravity-dominated Mixed Modes}",
      journal = {\apj},
         year = 2022,
        month = may,
       volume = {931},
       number = {1},
          eid = {64},
        pages = {64},
          doi = {10.3847/1538-4357/ac695b},
       adsurl = {https://ui.adsabs.harvard.edu/abs/2022ApJ...931...64Z}
}

@ARTICLE{2026ApJ..1002..198Z,
       author = {{Zhang}, Xinyi and {Li}, Tanda and {Chen}, Jianxing and {Chen}, Xinghao and {Li}, Gang and {Li}, Chunqian and {Bi}, Shaolan},
        title = "{Probing Red Giant Interiors with Gravity-dominated Mixed Modes. I. The Cases of KIC 9145955, KIC 9970396, KIC 9882316, and KIC 11968334}",
      journal = {\apj},
         year = 2026,
        month = may,
       volume = {1002},
       number = {2},
          eid = {198},
        pages = {198},
          doi = {10.3847/1538-4357/ae5e61},
archivePrefix = {arXiv},
       eprint = {2604.17382},
 primaryClass = {astro-ph.SR},
       adsurl = {https://ui.adsabs.harvard.edu/abs/2026ApJ..1002..198Z}
}

@ARTICLE{2025ApJS..279...37Z,
       author = {{Zhou}, Jianzhao and {Bi}, Shaolan and {Li}, Yaguang and {Yu}, Jie and {Li}, Tanda and {Zhang}, Xianfei and {Ye}, Lifei and {Li}, Mengjie and {Long}, Liu and {Sun}, Tiancheng and {Chen}, Yuqin},
        title = "{Asteroseismology of 687 TESS Red Giants: Individual Frequencies and Asymptotic Parameters}",
      journal = {\apjs},
         year = 2025,
        month = aug,
       volume = {279},
       number = {2},
          eid = {37},
        pages = {37},
          doi = {10.3847/1538-4365/adde57},
       adsurl = {https://ui.adsabs.harvard.edu/abs/2025ApJS..279...37Z}
}

@ARTICLE{2018ApJS..238...30Z,
       author = {{Zong}, Weikai and {Fu}, Jian-Ning and {De Cat}, Peter and {Shi}, Jianrong and {Luo}, Ali and {Zhang}, Haotong and {Frasca}, A. and {Corbally}, C.~J. and {Molenda-{\.Z}akowicz}, J. and {Catanzaro}, G. and {Gray}, R.~O. and {Wang}, Jiangtao and {Pan}, Yang and {Ren}, Anbing and {Zhang}, Ruyuan and {Jin}, Mengqi and {Wu}, Yue and {Dong}, Subo and {Xie}, Ji-Wei and {Zhang}, Wei and {Hou}, Yonghui and {LAMOST-Kepler Collaboration}},
        title = "{LAMOST Observations in the Kepler Field. II. Database of the Low-resolution Spectra from the Five-year Regular Survey}",
      journal = {\apjs},
         year = 2018,
        month = oct,
       volume = {238},
       number = {2},
          eid = {30},
        pages = {30},
          doi = {10.3847/1538-4365/aadf81},
archivePrefix = {arXiv},
       eprint = {1810.02753},
 primaryClass = {astro-ph.SR},
       adsurl = {https://ui.adsabs.harvard.edu/abs/2018ApJS..238...30Z}
}
\bibliographystyle{aasjournalv7}



\appendix
\restartappendixnumbering
\section{Internal Structures and Oscillation Frequencies}

\begin{figure}[!htbp]
\centering
\includegraphics[angle=0,scale=1.2]{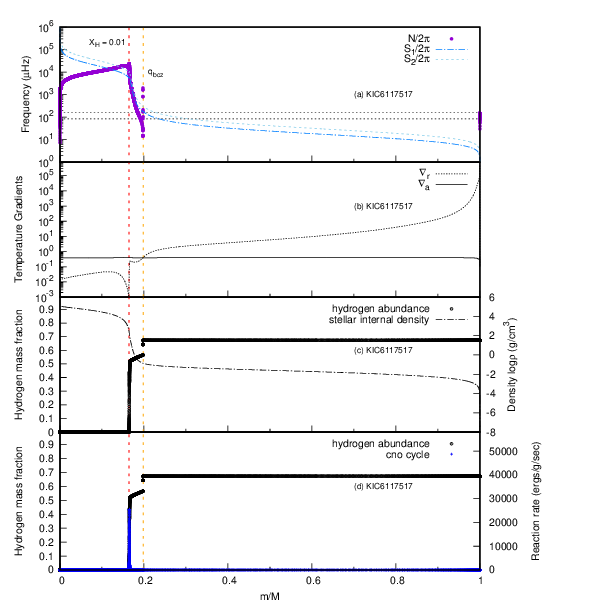}
\caption{Internal structure of KIC~6117517. The vertical red dashed lines in panel (a) to panel (d) denote the helium core boundary where the hydrogen mass fraction is about 0.01. The vertical orange dashed lines in panel (a) to panel (d) denote the base of the convection zone, which is marked as $q_{\rm bcz}$. In panel (a), two horizontal dashed lines represent the range of observed frequencies. The purple dot indicates the Brunt--V\"{a}is\"{a}l\"{a} frequency $N$, whereas the blue dash-dotted and light-blue dashed lines indicate the characteristic acoustic frequencies $S_1$ and $S_2$, respectively. In panel (b), the solid line and dotted line indicate adiabatic and radiative temperature gradients. In panel (c), circles and a dotted line indicate the hydrogen-abundance and density profiles. In panel (d), blue crosses represent the reaction rate of the carbon, nitrogen and oxygen (CNO) cycle.}
\label{fa1}
\end{figure}

\begin{figure}[!htbp]
\centering
\includegraphics[angle=0,scale=1.2]{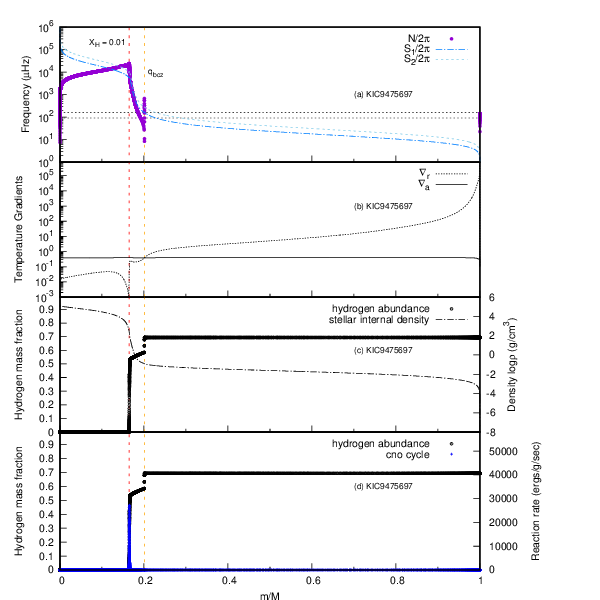}
\caption{Internal structure of KIC~9475697. The vertical red dashed lines in panel (a) to panel (d) denote the helium core boundary where the hydrogen mass fraction is about 0.01. The vertical orange dashed lines in panel (a) to panel (d) denote the base of the convection zone, which is marked as $q_{\rm bcz}$. In panel (a), two horizontal dashed lines represent the range of observed frequencies. The purple dot indicates the Brunt--V\"{a}is\"{a}l\"{a} frequency $N$, whereas the blue dash-dotted and light-blue dashed lines indicate the characteristic acoustic frequencies $S_1$ and $S_2$, respectively. In panel (b), the solid line and dotted line indicate adiabatic and radiative temperature gradients. In panel (c), circles and a dotted line indicate the hydrogen-abundance and density profiles. In panel (d), blue crosses represent the reaction rate of the carbon, nitrogen and oxygen (CNO) cycle.}
\label{fa2}
\end{figure}

\begin{figure}[!htbp]
\centering
\includegraphics[angle=0,scale=1.2]{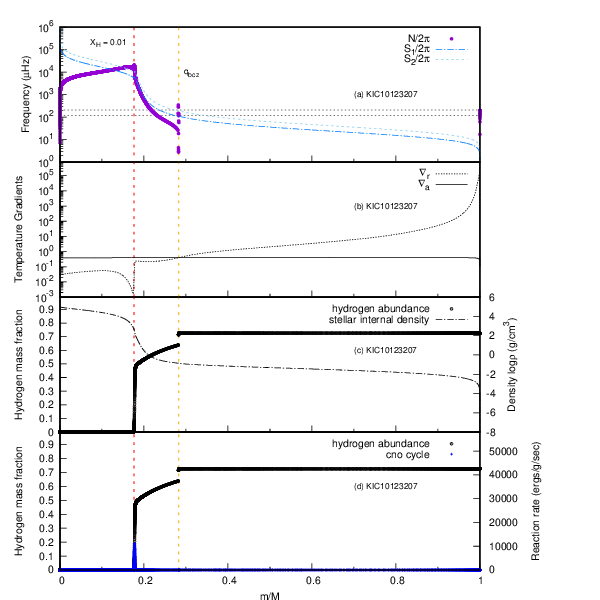}
\caption{Internal structure of KIC~10123207. The vertical red dashed lines in panel (a) to panel (d) denote the helium core boundary where the hydrogen mass fraction is about 0.01. The vertical orange dashed lines in panel (a) to panel (d) denote the base of the convection zone, which is marked as $q_{\rm bcz}$. In panel (a), two horizontal dashed lines represent the range of observed frequencies. The purple dot indicates the Brunt--V\"{a}is\"{a}l\"{a} frequency $N$, whereas the blue dash-dotted and light-blue dashed lines indicate the characteristic acoustic frequencies $S_1$ and $S_2$, respectively. In panel (b), the solid line and dotted line indicate adiabatic and radiative temperature gradients. In panel (c), circles and a dotted line indicate the hydrogen-abundance and density profiles. In panel (d), blue crosses represent the reaction rate of the carbon, nitrogen and oxygen (CNO) cycle.}
\label{fa3}
\end{figure}

\begin{figure}[!htbp]
\centering
\includegraphics[angle=0,scale=1.2]{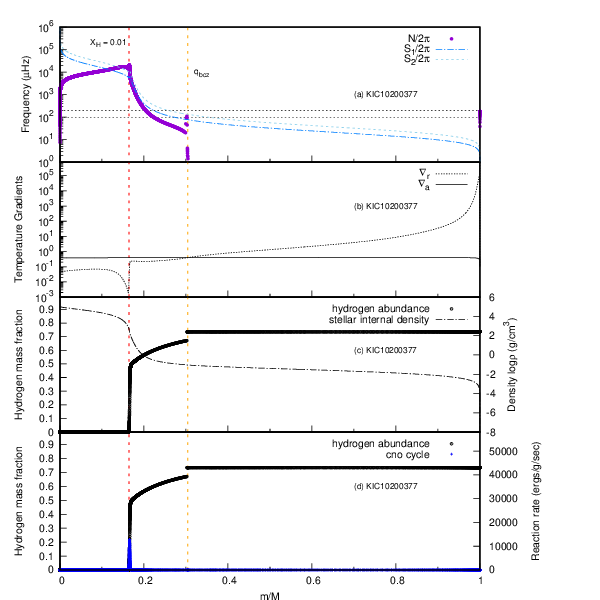}
\caption{Internal structure of KIC~10200377. The vertical red dashed lines in panel (a) to panel (d) denote the helium core boundary where the hydrogen mass fraction is about 0.01. The vertical orange dashed lines in panel (a) to panel (d) denote the base of the convection zone, which is marked as $q_{\rm bcz}$. In panel (a), two horizontal dashed lines represent the range of observed frequencies. The purple dot indicates the Brunt--V\"{a}is\"{a}l\"{a} frequency $N$, whereas the blue dash-dotted and light-blue dashed lines indicate the characteristic acoustic frequencies $S_1$ and $S_2$, respectively. In panel (b), the solid line and dotted line indicate adiabatic and radiative temperature gradients. In panel (c), circles and a dotted line indicate the hydrogen-abundance and density profiles. In panel (d), blue crosses represent the reaction rate of the carbon, nitrogen and oxygen (CNO) cycle.}
\label{fa4}
\end{figure}

This appendix contains four tables and four figures. Figures~\ref{fa1}--\ref{fa4} show the internal structures of the best-fitting models for KIC~6117517, KIC~9475697, KIC~10123207, and KIC~10200377, respectively. Panel (a) of each figure presents a propagation diagram showing the regions in which pressure and gravity waves propagate. Panel (b) shows the convective envelope; the convective boundary is identified by the Schwarzschild criterion, $\nabla_r=\nabla_a$. Panel (c) demonstrates that the stellar density profile exhibits a sharp change in slope at the boundary of the helium core. Panel (d) illustrates that the CNO-cycle reaction rate peaks near this boundary, coinciding with the location of the hydrogen-burning shell.

The observed frequencies and their uncertainties for KIC~6117517, KIC~9475697, KIC~10123207, and KIC~10200377 are listed in Tables~\ref{ta1}--\ref{ta4}, respectively. The p-dominated mixed modes identified in this work are marked in the last column.

\begin{table}[ht!]
\caption{\label{ta1}Oscillation frequencies of KIC 6117517 extracted by \citet{corsaro2015}. Frequency IDs are assigned in this work. The p-dominated mixed modes are marked in the last column.}
\centering
\begin{tabular}{cccccccccccc}
\hline\hline
ID    &$l$    &m  &Frequency  &Uncertainty   &Remark &ID          &$l$    &m  &Frequency  &Uncertainty   &Remark      \\
       &       &   &($\mu$Hz)  &($\mu$Hz)     &       &            &       &   &($\mu$Hz)  &($\mu$Hz)      &           \\
\hline
$f_{1}$	 &0	&0	&83.9104	&0.0107	&...               &$f_{35}$	&1	&-1	&118.8171	&0.0109	&...\\
$f_{2}$	 &0	&0	&93.7208	&0.0123	&...               &$f_{36}$	&1	&0	&119.0912	&0.0029	&p-dominated\\
$f_{3}$	 &0	&0	&103.4368	&0.0062	&...               &$f_{37}$	&1	&+1	&119.4202	&0.0100	&...\\
$f_{4}$	 &0	&0	&113.6025	&0.0021	&...               &$f_{38}$	&1	&0	&119.9339	&0.0002	&...\\
$f_{5}$	 &0	&0	&123.6357	&0.0031	&...               &$f_{39}$	&1	&0	&120.9997	&0.0011	&...\\
$f_{6}$	 &0	&0	&133.8028	&0.0080	&...               &$f_{40}$	&1	&0	&124.3937	&0.0008	&...\\
$f_{7}$	 &0	&0	&144.1375	&0.0267	&...               &$f_{41}$	&1	&0	&125.5703	&0.0004	&...\\
$f_{8}$	 &0	&0	&154.5449	&0.0297	&...               &$f_{42}$	&1	&0	&126.7652	&0.0004	&...\\
$f_{9}$	 &1	&0	&88.6524	&0.0060	&...               &$f_{43}$	&1	&0	&127.9242	&0.0004	&...\\
$f_{10}$ &1	&0	&89.0953	&0.0065 &p-dominated       &$f_{44}$	&1	&+1	&128.2715	&0.0009	&...\\
$f_{11}$ &1	&0	&89.5298	&0.0007 &...               &$f_{45}$	&1	&0	&128.8133	&0.0034	&p-dominated\\
$f_{12}$ &1	&0	&90.0822	&0.0025	&...               &$f_{46}$	&1	&0	&129.5676	&0.0034	&...\\
$f_{13}$ &1	&0	&90.6791	&0.0011	&...               &$f_{47}$	&1	&0	&130.7177	&0.0002	&...\\
$f_{14}$ &1	&0	&96.6666	&0.0011	&...               &$f_{48}$	&1	&0	&137.4327	&0.0005	&...\\
$f_{15}$ &1	&0	&97.3511	&0.0028	&...               &$f_{49}$	&1	&0	&138.6763	&0.0057	&...\\
$f_{16}$ &1	&0	&98.0453	&0.0004	&...               &$f_{50}$	&1	&0	&139.4766	&0.0074	&p-dominated\\
$f_{17}$ &1	&0	&98.5953	&0.0056	&p-dominated       &$f_{51}$	&1	&0	&140.6176	&0.0069	&...\\
$f_{18}$ &1	&0	&99.0693	&0.0010	&...               &$f_{52}$	&1	&0	&146.7663	&0.0012	&...\\
$f_{19}$ &1	&0	&99.7366	&0.0012	&...               &$f_{53}$	&1	&0	&148.3331	&0.0004	&...\\
$f_{20}$ &1	&0	&100.4944	&0.0022	&...               &$f_{54}$	&1	&0	&149.4613	&0.0153	&p-dominated\\
$f_{21}$ &1	&0	&106.1145	&0.0015	&...               &$f_{55}$	&1	&0	&150.5297	&0.0189	&...\\
$f_{22}$ &1	&0	&106.9760	&0.0006	&...               &$f_{56}$	&1	&0	&159.1129	&0.0189	&...\\
$f_{23}$ &1	&0	&107.8130	&0.0002	&...               &$f_{57}$	&1	&0	&160.1656	&0.0229	&p-dominated\\
$f_{24}$ &1	&0	&108.5235	&0.0034	&p-dominated       &$f_{58}$	&1	&0	&161.5668	&0.0158	&...\\
$f_{25}$ &1	&0	&109.0698	&0.0021	&...               &$f_{59}$	&2	&0	&82.3832	&0.0126	&...\\
$f_{26}$ &1	&0	&109.8274	&0.0004	&...               &$f_{60}$	&2	&0	&92.3720	&0.0064	&...\\
$f_{27}$ &1	&0	&110.7112	&0.0007	&...               &$f_{61}$	&2	&0	&102.1074	&0.0101	&...\\
$f_{28}$ &1	&0	&111.6417	&0.0009	&...               &$f_{62}$	&2	&0	&112.3045	&0.0058	&...\\
$f_{29}$ &1	&-1	&115.1371	&0.0016	&...               &$f_{63}$	&2	&0	&122.4213	&0.0041	&...\\
$f_{30}$ &1	&0	&115.5493	&0.0006	&...               &$f_{64}$	&2	&0	&132.5866	&0.0093	&...\\
$f_{31}$ &1	&0	&116.5621	&0.0005	&...               &$f_{65}$	&2	&0	&142.9067	&0.0214	&...\\
$f_{32}$ &1	&0	&117.5662	&0.0003	&...               &$f_{66}$	&2	&0	&153.3336	&0.0311	&...\\
$f_{33}$ &1	&-1	&118.1791	&0.0090	&...               &$f_{67}$	&2	&0	&163.6923	&0.0356	&...\\
$f_{34}$ &1	&0	&118.4574	&0.0027	&...               &...		   &...&...	&...		&...	&... \\
\hline
\end{tabular}
\end{table}

\begin{table}[ht!]
\caption{\label{ta2}Oscillation frequencies of KIC 9475697 extracted by \citet{corsaro2015}. Question marks in the $m$ column indicate azimuthal orders that could not be identified in \citet{corsaro2015}.} Frequency IDs are assigned in this work. The p-dominated mixed modes are marked in the last column.
\centering
\begin{tabular}{cccccccccccc}
\hline\hline
ID      &$l$    &$m$  &Frequency  &Uncertainty   &Remark &ID          &$l$    &$m$  &Frequency  &Uncertainty   &Remark      \\
        &       &     &($\mu$Hz)  &($\mu$Hz)     &       &            &       &     &($\mu$Hz)  &($\mu$Hz)      &           \\
\hline
$f_{1}$  & 0 & 0 & 81.5614  & 0.0294 & ...            & $f_{48}$ & 1 & 0    & 115.7320 & 0.0029 & p-dominated             \\
$f_{2}$  & 0 & 0 & 90.9681  & 0.0058 & ...            & $f_{49}$ & 1 & $+1$ & 115.9689 & 0.0004 & ...             \\
$f_{3}$  & 0 & 0 & 100.6017 & 0.0041 & ...            & $f_{50}$ & 1 & $-1$ & 116.1566 & 0.0002 & ...             \\
$f_{4}$  & 0 & 0 & 110.4707 & 0.0033 & ...            & $f_{51}$ & 1 & 0    & 116.4876 & 0.0006 & ...             \\
$f_{5}$  & 0 & 0 & 120.2252 & 0.0023 & ...            & $f_{52}$ & 1 & 0    & 117.4544 & 0.0004 & ...            \\
$f_{6}$  & 0 & 0 & 130.1600 & 0.0061 & ...            & $f_{53}$ & 1 & $+1$ & 117.8687 & 0.0005 & ...            \\
$f_{7}$  & 0 & 0 & 140.3016 & 0.0141 & ...            & $f_{54}$ & 1 & 0    & 118.4832 & 0.0004 & ...            \\
$f_{8}$  & 0 & 0 & 150.4119 & 0.0448 & ...            & $f_{55}$ & 1 & 0    & 119.5360 & 0.0010 & ...            \\
$f_{9}$  & 1 & 0 & 83.8974  & 0.0019 & ...            & $f_{56}$ & 1 & 0    & 120.6088 & 0.0014 & ...            \\
$f_{10}$ & 1 & 0 & 85.4863  & 0.0016 & ...            & $f_{57}$ & 1 & $-1$ & 121.2287 & 0.0006 & ...            \\
$f_{11}$ & 1 & 0 & 85.9867  & 0.0004 & ...            & $f_{58}$ & 1 & 0    & 121.6982 & 0.0002 & ...            \\
$f_{12}$ & 1 & 0 & 86.4125  & 0.0060 & ...            & $f_{59}$ & 1 & 0    & 122.8037 & 0.0016 & ...         \\
$f_{13}$ & 1 & 0 & 86.8321  & 0.0091 & ...            & $f_{60}$ & 1 & 0    & 123.9250 & 0.0003 & ...            \\
$f_{14}$ & 1 & 0 & 87.5257  & 0.0011 & ...            & $f_{61}$ & 1 & ?    & 124.4225 & 0.0168 & ...          \\
$f_{15}$ & 1 & 0 & 87.9898  & 0.0017 & ...            & $f_{62}$ & 1 & 0    & 124.9012 & 0.0029 & ...            \\
$f_{16}$ & 1 & 0 & 88.4230  & 0.0012 & ...            & $f_{63}$ & 1 & ?    & 125.2748 & 0.0031 & ...            \\
$f_{17}$ & 1 & 0 & 92.6725  & 0.0014 & ...            & $f_{64}$ & 1 & 0    & 125.5852 & 0.0039 & p-dominated    \\
$f_{18}$ & 1 & 0 & 93.3577  & 0.0018 & ...            & $f_{65}$ & 1 & $+1$ & 126.0170 & 0.0091 & ...          \\
$f_{19}$ & 1 & 0 & 95.2872  & 0.0033 & ...            & $f_{66}$ & 1 & 0    & 126.4989 & 0.0002 & ...            \\
$f_{20}$ & 1 & 0 & 95.8050  & 0.0068 &p-dominated     & $f_{67}$ & 1 & 0    & 127.6494 & 0.0035 & ...             \\
$f_{21}$ & 1 & 0 & 96.2681  & 0.0031 & ...            & $f_{68}$ & 1 & $-1$ & 133.3517 & 0.0111 & ...           \\
$f_{22}$ & 1 & 0 & 96.8716  & 0.0005 & ...            & $f_{69}$ & 1 & 0    & 133.9354 & 0.0039 & ...            \\
$f_{23}$ & 1 & 0 & 97.5508  & 0.0012 & ...            & $f_{70}$ & 1 & $+1$ & 134.5278 & 0.0098 & ...           \\
$f_{24}$ & 1 & 0 & 98.2560  & 0.0014 & ...            & $f_{71}$ & 1 & $-1$ & 134.9238 & 0.0199 & ...           \\
$f_{25}$ & 1 & 0 & 103.5840 & 0.0013 & ...            & $f_{72}$ & 1 & 0    & 135.0852 & 0.0063 & ...            \\
$f_{26}$ & 1 & 0 & 104.3720 & 0.0010 & ...            & $f_{73}$ & 1 & $+1$ & 135.3753 & 0.0206 & ...           \\
$f_{27}$ & 1 & 0 & 105.1212 & 0.0002 & ...            & $f_{74}$ & 1 & 0    & 135.8142 & 0.0064 & p-dominated    \\
$f_{28}$ & 1 & $+1$ & 105.2904 & 0.0013 & ...         & $f_{75}$ & 1 & ?    & 136.3724 & 0.0147 & ...           \\
$f_{29}$ & 1 & $-1$ & 105.6149 & 0.0115 &...          & $f_{76}$ & 1 & 0    & 136.9539 & 0.0047 & ...            \\
$f_{30}$ & 1 & 0 & 105.7226 & 0.0039    &p-dominated  & $f_{77}$ & 1 & 0    & 138.3644 & 0.0212 & ...           \\
$f_{31}$ & 1 & $+1$ & 105.8487 & 0.0009 &...          & $f_{78}$ & 1 & 0    & 144.1649 & 0.0021 & ...           \\
$f_{32}$ & 1 & $-1$ & 106.1042 & 0.0007 &...          & $f_{79}$ & 1 & $+1$ & 144.4964 & 0.0013 & ...                \\
$f_{33}$ & 1 & 0 & 106.2743 & 0.0030 &...             & $f_{80}$ & 1 & 0    & 145.2152 & 0.0250 & p-dominated             \\
$f_{34}$ & 1 & $-1$ & 106.8247 & 0.0013 &...          & $f_{81}$ & 1 & $+1$ & 145.5140 & 0.0345 & ...               \\
$f_{35}$ & 1 & 0 & 107.0336 & 0.0003 & ...            & $f_{82}$ & 1 & 0    & 146.1664 & 0.0174 & ...            \\
$f_{36}$ & 1 & 0 & 107.8765 & 0.0003 &...             & $f_{83}$ & 1 & $+1$ & 146.4116 & 0.0324 & ...           \\
$f_{37}$ & 1 & 0 & 111.4548 & 0.0013 &...             & $f_{84}$ & 1 & 0    & 154.1177 & 0.0269 & ...           \\
$f_{38}$ & 1 & $-1$ & 113.0133 & 0.0007 &...          & $f_{85}$ & 1 & 0    & 155.5508 & 0.0215 & p-dominated   \\
$f_{39}$ & 1 & 0 & 113.3324 & 0.0011 &...             & $f_{86}$ & 1 & 0    & 156.4901 & 0.0341 & ...           \\
$f_{40}$ & 1 & $+1$ & 113.6867 & 0.0007 &...          & $f_{87}$ & 2 & 0    & 80.1600  & 0.0225 & ...           \\
$f_{41}$ & 1 & $-1$ & 113.9444 & 0.0004 &...          & $f_{88}$ & 2 & 0    & 89.6403  & 0.0085 & ...            \\
$f_{42}$ & 1 & 0 & 114.2752 & 0.0003 &...             & $f_{89}$ & 2 & 0    & 99.2727  & 0.0051 & ...            \\
$f_{43}$ & 1 & $+1$ & 114.5938 & 0.0011 &...          & $f_{90}$ & 2 & 0    & 109.2126 & 0.0063 & ...            \\
$f_{44}$ & 1 & $-1$ & 114.8541 & 0.0007 &...          & $f_{91}$ & 2 & 0    & 119.0096 & 0.0036 & ...            \\
$f_{45}$ & 1 & 0 & 115.1256 & 0.0016 & ...            & $f_{92}$ & 2 & 0    & 128.9406 & 0.0092 & ...            \\
$f_{46}$ & 1 & $+1$ & 115.3931 & 0.0054 & ...         & $f_{93}$ & 2 & 0    & 139.1840 & 0.0181 & ...            \\
$f_{47}$ & 1 & $-1$ & 115.5532 & 0.0061 & ...         & $f_{94}$ & 2 & 0    & 149.1990 & 0.0439 & ...            \\
\hline
\end{tabular}
\end{table}

\begin{table}[ht!]
\caption{\label{ta3}Oscillation frequencies of KIC 10123207 extracted by \citet{corsaro2015}. Frequency IDs are assigned in this work. The p-dominated mixed modes are marked in the last column.}
\centering
\begin{tabular}{cccccccccccc}
\hline\hline
ID          &$l$&m  &Frequency  &Uncertainty   &Remark &ID          &$l$    &m  &Frequency  &Uncertainty   &Remark      \\
            &   &   &($\mu$Hz)  &($\mu$Hz)     &       &            &       &   &($\mu$Hz)  &($\mu$Hz)      &           \\
\hline
$f_{1}$		&0	&0	&127.0529	&0.0110	&...	        &$f_{23}$	&1	&0	&159.6858	&0.0022	&...   \\
$f_{2}$		&0	&0	&140.4660	&0.0065	&...	        &$f_{24}$	&1	&0	&161.0753	&0.0021	&p-dominated   \\
$f_{3}$		&0	&0	&154.1486	&0.0040	&...	        &$f_{25}$	&1	&0	&162.4891	&0.0020	&...   \\
$f_{4}$		&0	&0	&167.7083	&0.0031	&...	        &$f_{26}$	&1	&0	&169.0029	&0.0004	&...   \\
$f_{5}$		&0	&0	&181.5964	&0.0079	&...	        &$f_{27}$	&1	&0	&171.3331	&0.0003	&...   \\
$f_{6}$		&0	&0	&195.4977	&0.0330	&...	        &$f_{28}$	&1	&0	&173.5202	&0.0020	&...   \\
$f_{7}$		&1	&0	&119.6874	&0.0011	&...	        &$f_{29}$	&1	&0	&174.9683	&0.0042	&p-dominated   \\
$f_{8}$		&1	&0	&120.5964	&0.0086	&p-dominated	&$f_{30}$	&1	&0	&176.7934	&0.0029	&...   \\
$f_{9}$		&1	&0	&121.4157	&0.0036	&...	        &$f_{31}$	&1	&0	&184.5946	&0.0013	&...   \\
$f_{10}$	&1	&0	&122.4780	&0.0012	&...           	&$f_{32}$	&1	&0	&187.1407	&0.0057	&...   \\
$f_{11}$	&1	&0	&131.6569	&0.0021	&...           	&$f_{33}$	&1	&0	&188.8194	&0.0085	&p-dominated   \\
$f_{12}$	&1	&0	&132.9638	&0.0074	&...           	&$f_{34}$	&1	&0	&190.8828	&0.0063	&...   \\
$f_{13}$	&1	&0	&133.9665	&0.0033	&p-dominated	&$f_{35}$	&1	&0	&196.8273	&0.0074	&...   \\
$f_{14}$	&1	&0	&135.0076	&0.0004	&...           	&$f_{36}$	&1	&0	&199.9115	&0.0154	&...   \\
$f_{15}$	&1	&0	&136.3997	&0.0011	&...           	&$f_{37}$	&1	&0	&202.2877	&0.0228	&p-dominated   \\
$f_{16}$	&1	&0	&144.4224	&0.0005	&...           	&$f_{38}$	&1	&0	&204.0822	&0.0082	&...   \\
$f_{17}$	&1	&0	&146.0511	&0.0002	&...           	&$f_{39}$	&2	&0	&125.2673	&0.0135	&...   \\
$f_{18}$	&1	&0	&147.3360	&0.0020	&p-dominated	&$f_{40}$	&2	&0	&138.6456	&0.0073	&...   \\
$f_{19}$	&1	&0	&148.4595	&0.0001	&...           	&$f_{41}$	&2	&0	&152.2708	&0.0037	&...   \\
$f_{20}$	&1	&0	&150.0969	&0.0003	&...           	&$f_{42}$	&2	&0	&165.8629	&0.0049	&...   \\
$f_{21}$	&1	&0	&155.7870	&0.0003	&...           	&$f_{43}$	&2	&0	&179.7552	&0.0080	&...   \\
$f_{22}$	&1	&0	&157.7705	&0.0003	&...           	&$f_{44}$	&2	&0	&193.6605	&0.0223	&...   \\
\hline
\end{tabular}
\end{table}

\begin{table}[ht!]
\caption{\label{ta4}Oscillation frequencies of KIC 10200377 extracted by \citet{corsaro2015}. Frequency IDs are assigned in this work. The p-dominated mixed modes are marked in the last column.}
\centering
\begin{tabular}{cccccccccccc}
\hline\hline
ID      &$l$    &m  &Frequency  &Uncertainty   &Remark &ID          &$l$    &m  &Frequency  &Uncertainty   &Remark      \\
        &       &   &($\mu$Hz)  &($\mu$Hz)     &       &            &       &   &($\mu$Hz)  &($\mu$Hz)      &           \\
\hline
$f_{1}$		&0	&0	&91.4633	&0.0063	&...	        &$f_{28}$	&1	&0	&137.0296	&0.0003	&...    \\
$f_{2}$		&0	&0	&103.8863	&0.0146	&...	        &$f_{29}$	&1	&0	&143.2415	&0.0004	&...   \\
$f_{3}$		&0	&0	&115.8351	&0.0077	&...	        &$f_{30}$	&1	&0	&144.8624	&0.0002	&...   \\
$f_{4}$		&0	&0	&128.2301	&0.0047	&...	        &$f_{31}$	&1	&0	&146.3345	&0.0019	&...   \\
$f_{5}$		&0	&0	&140.6698	&0.0030	&...	        &$f_{32}$	&1	&0	&147.3654	&0.0033	&p-dominated   \\
$f_{6}$		&0	&0	&153.1591	&0.0035	&...	        &$f_{33}$	&1	&0	&148.6954	&0.0002	&...   \\
$f_{7}$		&0	&0	&165.8603	&0.0096	&...	        &$f_{34}$	&1	&0	&154.0643	&0.0004	&...   \\
$f_{8}$		&0	&0	&178.7340	&0.0262	&...	        &$f_{35}$	&1	&0	&155.9684	&0.0003	&...   \\
$f_{9}$		&0	&0	&191.4935	&0.0612	&...	        &$f_{36}$	&1	&0	&157.8311	&0.0058	&...   \\
$f_{10}$	&1	&0	&107.9525	&0.0017	&...	        &$f_{37}$	&1	&0	&159.3519	&0.0048	&p-dominated   \\
$f_{11}$	&1	&0	&108.8450	&0.0042	&...	        &$f_{38}$	&1	&0	&160.5416	&0.0033	&...   \\
$f_{12}$	&1	&0	&109.6778	&0.0051	&...	        &$f_{39}$	&1	&0	&168.8175	&0.0027	&...   \\
$f_{13}$	&1	&0	&110.3390	&0.0073	&p-dominated	&$f_{40}$	&1	&0	&170.9541	&0.0064	&   \\
$f_{14}$	&1	&0	&111.1040	&0.0165	&...	        &$f_{41}$	&1	&0	&172.4400	&0.0103	&p-dominated   \\
$f_{15}$	&1	&0	&112.0248	&0.0083	&...	        &$f_{42}$	&1	&0	&174.0295	&0.0094	&...   \\
$f_{16}$	&1	&0	&119.5208	&0.0013	&...	        &$f_{43}$	&1	&0	&181.2867	&0.0014	&...   \\
$f_{17}$	&1	&0	&120.6355	&0.0005	&...	        &$f_{44}$	&1	&0	&183.7600	&0.0308	&...   \\
$f_{18}$	&1	&0	&121.6450	&0.0044	&...	        &$f_{45}$	&1	&0	&185.3816	&0.0193	&p-dominated   \\
$f_{19}$	&1	&0	&122.4300	&0.0027	&p-dominated	&$f_{46}$	&2	&0	&102.1328	&0.0110	&   \\
$f_{20}$	&1	&0	&123.3749	&0.0003	&...	        &$f_{47}$	&2	&0	&114.1223	&0.0180	&...   \\
$f_{21}$	&1	&0	&124.5287	&0.0012	&...	        &$f_{48}$	&2	&0	&126.5206	&0.0040	&...   \\
$f_{22}$	&1	&0	&129.6672	&0.0010	&...	        &$f_{49}$	&2	&0	&139.0093	&0.0047	&...   \\
$f_{23}$	&1	&0	&131.0234	&0.0005	&...	        &$f_{50}$	&2	&0	&151.4301	&0.0085	&...   \\
$f_{24}$	&1	&0	&132.3868	&0.0003	&...	        &$f_{51}$	&2	&0	&164.2079	&0.0161	&...   \\
$f_{25}$	&1	&0	&133.6857	&0.0001	&...	        &$f_{52}$	&2	&0	&177.1014	&0.0241	&...   \\
$f_{26}$	&1	&0	&134.6824	&0.0026	&p-dominated	&$f_{53}$	&2	&0	&189.7988	&0.0338	&   \\
$f_{27}$	&1	&0	&135.6766	&0.0002	&...	        &...  &...   &...	&...	        &...       &... \\

\hline
\end{tabular}
\end{table}

\end{CJK*}
\end{document}